\documentclass[a4paper,11pt]{article}
\usepackage[margin=2.5cm]{geometry}
\usepackage[utf8]{inputenc}
\usepackage[T1]{fontenc}
\usepackage{hyperref}
\usepackage{amsmath,amsfonts,esdiff,nicefrac,amssymb}
\usepackage{tikz}
\usepackage{caption}
\usepackage{authblk}

\graphicspath{{figpdf/}}

\makeatletter

\@addtoreset{equation}{section}
\makeatother
\newcommand{\secref}[1]{Sec.~\ref{sec:#1}}

\newcommand{\normal}[1]{\ :\! #1 \!\!:}
\newcommand{\aver}[1]{\left\langle {#1} \right\rangle}
\newcommand{\bra}[1]{\langle {#1} |}

\newcommand{\vect}[1]{\boldsymbol{#1}}

\newcommand{\id}{\mathbf{1}}

\newcommand{\nn}{\nonumber}
\newcommand{\wh}{\widehat}
\newcommand{\wt}{\widetilde}

\newcommand{\dif}[1]{\ensuremath{\operatorname{d}\!{#1}}}

\newmuskip\pFqskip
\mathchardef\pFcomma=\mathcode`, 
\newcommand*\pFq[5]{%
  \begingroup
  \begingroup\lccode`~=`,
    \lowercase{\endgroup\def~}{\pFcomma\mkern\pFqskip}%
  \mathcode`,=\string"8000
  {}_{#1}\mathrm{F}_{#2}\biggl[\genfrac..{0pt}{}{#3}{#4};#5\biggr]%
  \endgroup
}

               {Remark:
                 \begin{list}{}%
                   {\setlength{\leftmargin}{0.07\linewidth}}%
                 \item[]%
                   \footnotesize
               }
               {\end{list}}

\title{The imaginary Toda field theory}
\author[]{T. Dupic}
\author[]{B. Estienne}
\author[]{Y. Ikhlef}
\affil[]{Sorbonne Universit\'e, CNRS, Laboratoire de Physique Th\'eorique et Hautes \'Energies, LPTHE, F-75005 Paris, France}

\date{}

\begin{document}

\maketitle

\begin{abstract}
We consider the two-dimensional $\mathfrak{sl}_n$ quantum Toda field theory with an imaginary background charge. This conformal field theory has a higher spin symmetry ($W_n$ algebra), a central charge $c \leq n-1$ and a continuous spectrum. Using the conformal bootstrap, we compute structure constants involving two arbitrary scalar fields and a semi-degenerate field of Wyllard type. The solution obtained is not the analytic continuation of the usual Toda three-point function. Non-scalar primary fields and their three-point functions are also discussed. Non-scalar primary fields are classified by conjugacy classes of the permutation group $\mathfrak{S}_n$, and their structure constants are computed explicitly, up to an overall factor. 
\end{abstract}

\section{Introduction}
\label{sec:intro}

In the context of Conformal Field Theory (CFT) applied to critical models of Statistical Mechanics, the understanding of the operator algebra, {\it i.e.} the spectrum of primary operators, their fusion rules and the structure constants appearing in the Operator Product Expansion (OPE), is of central importance. While the conformal dimensions determine the critical exponents of a given universality class, the structure constants (together with the conformal blocks) are the building blocks for $N$-point correlation functions on the sphere.

The Liouville theory with an imaginary background charge \cite{Schomerus03,KostovPetkova06,Zamo05} is an example of non-rational CFT with a central charge $c\leq 1$ and a continuous spectrum. The structure constants of the imaginary Liouville theory have been obtained in  \cite{Zamo05} using the analytic conformal bootstrap. A striking result is that they are {\it not} the analytic continuation of the usual Liouville structure constants (with a real background charge).  In \cite{centralchargeless1} the spectrum was argued to be spacelike, and crossing symmetry was checked numerically. 

Critical models (Ising, RSOS, Potts, percolation, O($n$) loop models, {\it etc.}) typically have a discrete (finite or infinite) spectrum. Given that the imaginary Liouville theory has a continuous spectrum of primary operators, it can be rather surprising that this theory has applications in the context of statistical physics.  Yet it has been observed numerically that the imaginary Liouville theory is relevant for the study of certain three point functions for percolation\cite{DelfinoViti11}, and more generally for critical cluster and loop models \cite{PSVD13,IJS16}. This apparent contradiction is alleviated upon noticing that the lattice observables concerned are typically non-local, such as the probability for three points to lie on the same loop or cluster. So one is in fact dealing with non-rational CFTs, in all likelihood logarithmic \cite{1742-5468-2007-09-P09002,EI15}. 

Similar questions can be asked for models with extended symmetries. In particular many critical integrable lattice models -- vertex and face models -- fall into the universality class of  $W_n$-symmetric CFTs~\cite{Pasquier-sln}, and their central charge is $c \leq n-1$. Likewise the Fully-Packed Loop (FPL) model on the honeycomb lattice~\cite{reshetikhin91,Kondev-FPL} displays some features of a $W_3$ symmetry~\cite{DEI-FPL}. In this context a natural question to ask is about the relevance of the $\mathfrak{sl}_n$ Toda field theory with an imaginary background charge. The $W_n$ analog of the Dorn-Otto-Zamolodchikov-Zamolodchikov (DOZZ) formula \cite{DornOtto92,ZZ96}, {\it i.e.} the structure constant of three vertex operators for the $\mathfrak{sl}_n$ Toda field theory with a real background charge $Q$, was partially obtained by the conformal bootstrap procedure in \cite{FL1,FL2}. The results are restricted to the case when one of the operators is ``semi-degenerate'' with respect to the $W_n$ algebra. We compute these structure constants in the imaginary $\mathfrak{sl}_n$ Toda CFT. The case $n=4$ was already studied in \cite{Furlan:2015ska}. Similarly to the case of Liouville, these structure constants are not the analytic continuation of the usual Toda CFT : their analytic expressions for real and imaginary $Q$ are different.

In many physical situations (such as loop models, non-diagonal minimal models) one has to deal with non-scalar primary fields, \emph{i.e.} fields with a nonzero conformal spin. In the case of CFTs with only a Virasoro algebra, crossing symmetry leads to severe constraints on the possible non-scalar fields \cite{EI15,Migliaccio2017}. We reconsider this question for CFTs boasting a $W_n$ symmetry.  We classify the possible non-scalar primary fields under the assumption of well-defined monodromies with fully-degenerate fields. Such non-scalar fields are found to be classified by conjugacy classes of the permutation group $\mathfrak{S}_n$, and in particular scalar fields correspond to the trivial permutation. Finally we consider the analytic bootstrap in the presence of these non-scalar operators: we obtain two families of shift equations for them, which can be solved explicitly when one of the operators is scalar.

\section{General background}
\label{sec:background}

A complete construction of $W_n$ symmetry algebras and their representation theories can be found in \cite{FATEEV1988,bouwknegt1993w}. In this section we recall some basic facts and notations, starting with $W_3$.

\subsection{The \texorpdfstring{$W_3$}{W3} algebra}
The $W_3$ algebra is a chiral symmetry algebra generated by two fields \cite{Zamolodchikov85,Fateev87}: the spin-$2$ stress energy tensor, ensuring conformal invariance,  and an additional spin-$3$ current $W(z)$. The operator algebra generated by these two operators is completely defined by the Operator Product Expansion (OPE):

\begin{equation}
  \label{eq:OPE_W_T}
  \begin{split}
    T(z) T(0) &= \frac{c}{2 z^4} + \frac{2}{z^2} T(0) + \frac{1}{z} \partial T(0)
    + \text{reg,} \\
    T(z) W(0) &= \frac{3}{z^2} W(0) + \frac{1}{z} \partial W(0) + \text{reg,} \\
    W(z) W(0) &= \frac{c}{3 z^6} + \frac{2 T(0)}{z^4} + \frac{\partial T(0)}{z^3} + \frac{1}{z^2} \left[\frac{3}{10} \partial^2 T(0) + \frac{32}{22+ 5 c} \Lambda(0) \right] \\ & + \frac{1}{z} \left[\frac{1}{15} \partial^3 T(0) + \frac{16}{22+5 c} \partial\Lambda(0) \right] + \text{reg,}
  \end{split}
\end{equation}
where $\Lambda$ is the (quasi-primary) composite field $\Lambda(z) = \normal{T^2(z)} - \frac{3}{10} \partial^2 T(z)$.
The first relation defines the Virasoro algebra (central charge $c$), the second one expresses that $W$ is a primary field (\emph{w.r.t.} the Virasoro algebra) of dimension $3$, while the third one comes from the closure of the operator algebra.

The modes of these two operators are defined by:
\[
L_{n} \Phi(\xi) = \oint \dif u (u - \xi)^{n+1} T(u) \Phi(\xi),
\qquad W_{n} \Phi(\xi) = \oint \dif u (u - \xi)^{n+2} W(u) \Phi(\xi)
\]
The OPEs defined previously are equivalent to the following commutation relations between the modes:
\begin{equation}
  \label{eq:W3_relations}
  \begin{split}
    \left[L_n, L_m \right] &= (n-m) L_{n+m} + \frac{c}{12}(n^3 - n) \delta_{n+m,0} \\
    \left[L_n, W_m \right] &= (2n - m) W_{n+m} \\
    \left[W_n, W_m \right] &= \frac{c}{360} (n^2 - 4) (n^2 - 1) n \delta_{n+m,0} + \frac{16}{22+ 5 c} (n-m) \Lambda_{n+m} \\  
    & \quad +  \frac{1}{30}(n-m) \left( 2m^2 + 2 n^2 - mn - 8 \right) L_{n+m} \,,
  \end{split}
\end{equation}
where the $\Lambda_n$ are given by 
\begin{align}
  \Lambda_n = \sum_{k < n/2} L_{k} L_{n-k} + \sum_{k \geq n/2} L_{n-k} L_{k}  + \frac{\left(1 + \left\lceil \frac{n}{2} \right \rceil \right)\left(1 + \left\lfloor \frac{n}{2} \right \rfloor \right)}{5} L_n.
\end{align}

A primary field for $W_3$ is a field $\Phi$ which is primary with respect to both operators: 
\[
\forall n>0 \,, \qquad W_n \Phi = L_n \Phi = 0 \,,
\]
and
\[
W_{0} \Phi = w_{\Phi} \Phi, \qquad L_{0} \Phi = \Delta_{\Phi} \Phi \,,
\]
$\Delta_{\Phi}$ is the conformal dimension of the field.  $(\Delta_{\Phi}, w_{\Phi})$ identifies a field in a $W_3$-conformal theory (we assume that there are no degeneracies in the spectrum). Based on those conditions, the OPE of the fields $T$ and $W$ with the primary fields can be written:
\begin{equation}
  \begin{split}
    T(z)\Phi(0) &= \frac{\Delta_\Phi \Phi(0)}{z^2} + \frac{\partial \Phi(0)}{z}  + \text{reg.} \\
    W(z)\Phi(0) &= \frac{w_{\Phi} \Phi(0)}{z^3} + \frac{W_{-1} \Phi(0)}{z^2} + \frac{W_{-2} \Phi(0)}{z} + \ \text{reg.} 
  \end{split}
\end{equation}
As a consequence, $T(z)$ and $W(z)$ behave as $\frac{1}{z^4}$ and $ \frac{1}{z^6}$ respectively for $ z \rightarrow \infty$. In particular global $W_3$ symmetry fixes the two point functions to be of the form
\[
\langle \Phi_1(z_1) \Phi_2(z_2) \rangle = \frac{C_{12}}{ \left( z_{1}-z_2 \right)^{2 \Delta_1}} \, \delta_{\Delta_1,\Delta_2}\, \delta_{w_1+w_2,0} \,.
\]
The conjugate $\Phi_i^{*}$ of the primary field $\Phi_i$ is the unique primary field such that  
\begin{align*}
\langle \Phi_i^*(z_1) \Phi_j(z_2) \rangle = \frac{\delta_{ij}}{ \left( z_{1}-z_2 \right)^{2 \Delta_{i}}}  \,.
\end{align*}
In view of the previous relation, $w_i = -w_{j^*}$ and  $\Delta_{i} = \Delta_{j^*}$.  In order to ensure $W_0 = W_0^{\dag}$ one is led to define
\begin{align*}
\bra{\Phi_i} = \lim_{z \to \infty} z^{2 \Delta_{i}}  \bra{0} \Phi_i^*(z) \,. 
\end{align*}
If $\Phi_1$ and $\Phi_2$ are two primary fields \emph{w.r.t.} the $W_3$ algebra and the fusion $\Phi_1 \times \Phi_2 \rightarrow \Phi_p + \cdots$ is allowed, then the OPE of $\Phi_1$ with $\Phi_2$ will also involve all the descendants of $\Phi_{p}$, with respect to $T$ and $W$,
\[
\Phi_1(z) \Phi_2(0) = \sum_{\vect{n}, \vect{m}} z^{|\vect{n}|+|\vect{m}| + \Delta_p - \Delta_1 - \Delta_2}C_{\Phi_1, \Phi_2}^{W_{-\vect{n}} L_{-\vect{m}} \Phi_p} W_{-\vect{n}} L_{-\vect{m}} \Phi_p(0), 
\]
where, for $\vect{n}$ a multi-indices vector $X_{-\vect{n}} = X_{-n_1} X_{-n_2} \cdots$ and $|\vect{n}| = n_1 + n_2 + \cdots$. 

Contrary to the $L_{n}$ operators, the modes $W_n$ do not act on correlation functions as differential operators.  While every three-point function involving Virasoro descendant operators can be expressed in terms of three-point functions of primaries, it is not the case for $W_3$ descendants:some of the coefficients $C_{\Phi_1, \Phi_2}^{W_{-\vect{n}} L_{-\vect{m}} \Phi_p}$ in the OPE above cannot be expressed in terms of $C_{\Phi_1, \Phi_2}^{\Phi_p}$ just by using the commutation relations \eqref{eq:W3_relations}.

\subsection{Parametrisation of $W_3$ conformal weights}

Many features of the $W_3$ Toda field theory are related to the $\mathfrak{sl}_3$ Lie algebra (see Appendix~\ref{app:sln_conventions} for notations). We parameterise the central charge in terms of a background charge $\vect{Q}$
\begin{equation} \label{eq:cc}
  c = 2 - 12 \vect{Q}\cdot \vect{Q} \,,
  \qquad \vect{Q} = \left( \frac{1}{b} - b \right) \, \vect\rho \,,
\end{equation}
where $\vect{\rho}$ is the Weyl vector of $\mathfrak{sl}_3$.
The conformal weights $(\Delta, w)$ (the eigenvalues of $L_0$ and $W_0$, respectively) are expressed in terms of a two-dimensional ``charge'' vector $\vect\alpha$ as follows
\begin{equation} \label{eq:Delta-w}
  \begin{aligned}
    \Delta_{\vect\alpha} &= \frac{1}{2} \vect{\alpha}\cdot \left(\vect{\alpha} - 2 \vect{Q} \right) \,, \\
    w_{\vect\alpha} &= \sqrt\frac{48}{22+5 c} \left(\vect{\alpha} - \vect{Q}\right) \cdot\vect{h}_1 \times \, \left(\vect{\alpha} - \vect{Q}\right) \cdot \vect{h}_2 \times \left(\vect{\alpha} - \vect{Q}\right) \cdot \vect{h}_3 \,.
  \end{aligned}
\end{equation}
We will also use the notation $\vect{P} = \vect{\alpha} - \vect{Q}$.

There is a natural action of the Weyl group $\mathcal{W} \simeq \mathfrak{S}_3$ on charge vectors:
\begin{equation} 
  \vect{\alpha} \to \sigma \star \vect{\alpha} = \vect{Q}+ \sigma(\vect\alpha-\vect{Q}) \,,
\end{equation}
under which the weights $\Delta_{\vect\alpha}$ and $w_{\vect\alpha}$ are invariant
\begin{equation}
  \label{Weyl transformations}
  \forall \sigma \in \mathcal{W} \,,
  \qquad (\Delta_{\sigma \star \vect{\alpha}}, w_{\sigma \star \vect{\alpha}})
  = (\Delta_{\vect\alpha}, w_{\vect\alpha}) \,.
\end{equation}
For now we only reason in terms of representation theory of the $W_3$ algebra, and we do not specify any local Lagrangian defining the correlation functions. In this context, we denote by $\Phi_{\vect\alpha}(z)$ a primary field with weights $(\Delta_{\vect\alpha}, w_{\vect\alpha})$. Therefore, we identify 
\begin{equation} \label{eq:relabel}
  \forall \sigma \in \mathcal{W} \,,
  \qquad \Phi_{\vect\alpha} \equiv \Phi_{\sigma \star \vect\alpha} \,.
\end{equation}
Finally, the dual of the field $\Phi_{\vect\alpha}$ is $\Phi_{\vect\alpha^*}$ (see Appendix \ref{app:sln_conventions} for the definition of $\vect\alpha^*$) as can be seen from  
\begin{align*}
  (\Delta_{\vect\alpha^*}, w_{\vect\alpha^*}) =  (\Delta_{\vect\alpha}, -w_{\vect\alpha}) \,.
\end{align*}
Note that $2 \vect Q - \vect \alpha$ and $\vect\alpha^*$ are equivalent since $2 \vect Q - \vect \alpha = s_3 \star \vect{\alpha}^{*}$.

\subsection{Semi-degenerate and fully-degenerate representations of $W_3$}
\label{sec:degenerate-rep}

Some particular $\mathcal{W}_3$ primary fields have null vectors (i.e. some of their descendants are linearly dependent). These primary fields are called degenerate. Two types of degenerate representations must be distinguished
\begin{itemize}
\item semi-degenerate representations, satisfying only one null-vector equation. The conformal dimension and the $W$-charge of the semi-degenerate primaries are related by a polynomial equation.
\item fully degenerate representations, with two or more null-vector equations, for which both the conformal dimension and the $W$-charge are fixed (for a given central charge).
\end{itemize}

For semi-degenerate representations, there is only one primitive linear relation between descendants.  The simplest example is the representation generated by an operator with a null vector at level one
\begin{equation}
  L_{-1} \Phi_{\vect\alpha} \propto W_{-1} \Phi_{\vect\alpha},
  \label{eq:mode_semi_degenerate}
\end{equation}
which requires the relation
\begin{equation}
  9 w_{\vect\alpha}^2 = \frac{2\Delta_{\vect\alpha}^2 \left(32 \Delta_{\vect\alpha} + 2 - c\right)}{22 + 5 c} \,.
  \label{eq:poly_semi_degenerate}
\end{equation}
In terms of charge vectors, this means  (up to Weyl transformations)
\begin{align}
  \vect{\alpha} = \kappa \vect{\omega}_1
  \qquad \text{or} \qquad \vect{\alpha} = \kappa \vect{\omega}_2 \,,
\end{align}
where $\kappa \in \mathbb{C}$. The explicit relation between descendants is:
\begin{equation}
  \label{eq:W3_semi_degenerate_nve}
  W_{-1} \Phi_{\vect\alpha} = \frac{3 w}{2 \Delta} L_{-1} \Phi_{\vect\alpha} \,.
\end{equation}

Fully degenerate states possess additional null vectors. Focusing on the simplest type of degenerate representation, consider a field $\Phi$ satisfies the semi-degenerate condition \eqref{eq:W3_semi_degenerate_nve}, together with a linear relation at level two, between $\{W_{-2}\Phi_{\vect\alpha}, L_{-2}\Phi_{\vect\alpha}, L^2_{-1}\Phi_{\vect\alpha}\}$. The algebra~\eqref{eq:W3_relations} then yields four possible values $\vect\alpha \in \{ b \vect{\omega}_1, b \vect{\omega}_2, -\vect{\omega}_1/b, -\vect{\omega}_2/b \}$, where $b$ is related to the background charge $\vect Q$ by~\eqref{eq:cc}. The precise form of the additional null-vector conditions is then (see \cite{Bowcock:1992gt}):
\begin{equation}
  \label{eq:null_vector_W3}
  \begin{split}
& \left( W_{-2} - \frac{12 w}{\Delta(5 \Delta + 1)}L_{-1}^2 + \frac{6 w (\Delta + 1)}{\Delta (5 \Delta + 1)} L_{-2} \right) \Phi_{\vect\alpha} = 0\,,  \\ 
& \left( W_{-3} - \frac{16 w}{\Delta(\Delta + 1)(5\Delta + 1)} L_{-1}^3 + \frac{12 w}{\Delta (5\Delta + 1)} L_{-1}L_{-2} + \frac{3 w}{2 \Delta} \frac{(\Delta - 3)}{(5\Delta + 1)} L_{-3} \right) \Phi_{\vect\alpha} = 0 \,,
\end{split}
\end{equation}
where $\Delta$ and $w$ are given by~\eqref{eq:Delta-w}. 

\subsection{Fusion in $W_3$}

The above null-vector equations impose constraints on the fusion rules and on the structure constants involving degenerate fields. In particular, we can extract from them the fusion rules of the fully degenerate states. For example, one can show that the correlation function $\langle \Phi_{\vect{\alpha}}\Phi_{b\vect\omega_1}  \Phi_{\vect\beta^{*}} \rangle$ vanishes unless $\vect\beta = \vect\alpha + b \vect{h}_j$ for some $j \in \{1,2,3\}$. Fusion rules match the behaviour of tensor product of representation in $\mathfrak{su}(3)$: 
\begin{equation}
  \Phi_{b\vect{\omega}_1} \otimes \Phi_{\vect\alpha} \rightarrow  \bigoplus_{j=1}^3 \Phi_{\vect\alpha + b \vect{h}_j}, \qquad     \Phi_{b\vect{\omega}_2} \otimes \Phi_{\vect\alpha} \rightarrow \bigoplus_{j=1}^3 \Phi_{\vect\alpha - b \vect{h}_j} \, .\label{eq:fusion_rule_degenerate}
\end{equation}
This result can be found in \cite{FL1}, it can also be extracted from the result in the Appendix~\ref{app:hypergeometric_sl3}, by setting $\kappa = 0$ in the differential equation. Note however that generically there will be nontrivial multiplicities in these fusion rules. This is why a generic four-point function involving a fully degenerate field $  \Phi_{b\vect{\omega}_1}$ does not obey a differential equation of order three. If one of the other fields is semi-degenerate though, a differential equation can be found, but its order depends on the semi-degenerate field \cite{Belavin:2016qaa,Belavin2017}.

For semi-degenerate operators, similar (if less strict) restrictions exist. Let $\Phi_1$ and $\Phi_2$ be two semi-degenerate fields at level $1$. The residue theorem applied to the contour integral
$$ \oint_C dz \,  (z-1)^2 z \, \langle \Phi_{2} | W(z) \Phi_{\vect{\alpha}}(1) | \Phi_{1} \rangle \,,$$
where $C$ is a contour enclosing $0$ and $1$, yields
\begin{equation*}
 \langle \Phi_{2} | W_1 \Phi_{\vect{\alpha}}(1) | \Phi_{1}  \rangle  -  \ \langle  \Phi_{2} | \Phi_{\vect{\alpha}}(1) W_{-1} |  \Phi_{1}  \rangle + \left( 2(w_1 - w_{2}) - w_{\vect{\alpha}} \right) \langle \Phi_2 | \Phi_{\vect{\alpha}}(1) | \Phi_1  \rangle =0 \,.
\end{equation*}
Now using the null-vector condition $W_{-1} \Phi_j = \frac{3 w_j}{2 \Delta_j} L_{-1}\Phi_j$, and 
\begin{align*}
\langle \Phi_2 | \Phi_{\vect{\alpha}}(1) L_{-1} | \Phi_1 \rangle & =( \Delta_{1} + \Delta_{\vect{\alpha}} - \Delta_2) \langle \Phi_2 | \Phi_{\vect{\alpha}}(1) | \Phi_1 \rangle , \\
 \langle \Phi_2 | L_1 \Phi_{\vect{\alpha}}(1)| \Phi_1 \rangle & =( \Delta_{2} + \Delta_{\vect{\alpha}} - \Delta_1) \langle \Phi_2 | \Phi_{\vect{\alpha}}(1) | \Phi_1 \rangle 
\end{align*}
we get the condition for $\langle\Phi_2|\Phi_{\vect\alpha}(1)|\Phi_1 \rangle$ to be nonzero:
\begin{equation*}
w_{\vect{\alpha}}  = \frac{3 w_2}{2 \Delta_2} ( \Delta_{2} + \Delta_{\vect{\alpha}} - \Delta_1)   -  \frac{3 w_1}{2 \Delta_1}( \Delta_{1} + \Delta_{\vect{\alpha}} - \Delta_2) + 2(w_1 - w_{2}) \,.
\end{equation*}
This restricts the fusion rules (up to Weyl transformations) as follows:
\begin{equation}
  \label{eq:fusion_rules_semi_degenerate}
  \begin{split}
    \aver{\Phi_{\kappa \vect\omega_1} \Phi_{\kappa' \vect\omega_1} \Phi_{\beta^*}} &\neq 0
    \quad \text{for} \quad \beta \in \mathbb{R}  \vect{e}_1 + (\kappa + \kappa') \vect{\omega}_{1} \,, \\
    \aver{\Phi_{\kappa \vect\omega_2} \Phi_{\kappa' \vect\omega_2} \Phi_{\beta^*}} &\neq 0
    \quad \text{for} \quad \beta \in \mathbb{R}  \vect{e}_2 + (\kappa + \kappa') \vect{\omega}_{2}  \,, \\
    \aver{\Phi_{\kappa \vect\omega_1} \Phi_{\kappa' \vect\omega_2} \Phi_{\beta^*}} &\neq 0
    \quad \text{for} \quad \beta \in \mathbb{R} \vect{\rho} + \frac{\kappa' - \kappa}{2} \vect{h}_{2} \,.
  \end{split}
\end{equation}
A special case is the fusion between a semi- and a fully degenerate field. It is possible to explicitly describe the allowed operators in the fusion $\Phi_{b \vect{\omega}_1} \otimes \Phi_{\kappa \vect{\omega}_i} $. Putting together the previous fusion rules we get
\begin{equation}
  \label{eq:fusion_rules_degenerate_semi_degenerate}
  \begin{split}
    & \Phi_{b \vect\omega_1} \otimes  \Phi_{\kappa \vect\omega_1} \rightarrow
    \Phi_{\kappa \vect\omega_1 + b \vect{h}_1 } \oplus \Phi_{\kappa \vect\omega_1 + b \vect{h}_2} \,, \\
    & \Phi_{b \vect\omega_1} \otimes  \Phi_{\kappa \vect\omega_2} \rightarrow
    \Phi_{\kappa \vect\omega_2 + b \vect{h}_1 } \oplus \Phi_{\kappa \vect\omega_2 + b \vect{h}_3} \,. 
  \end{split}
\end{equation}

The semi- and fully degenerate representations of higher level can be described explicitly by studying level-$N$ null vectors. The fields $\Phi_{\kappa \vect\omega_2 + b \vect h_1}$ and $\Phi_{\kappa \vect\omega_2 + b \vect h_3}$, which appear in the previous fusions are also semi-degenerate, see for example \cite{Belavin:2016qaa} for an explicit derivation.

\subsection{\texorpdfstring{$W_n$}{Wn} algebra and conventions for \texorpdfstring{$\mathfrak{sl}_n$}{sl(n)}}

The $W_n$ algebra is a generalisation of the $W_3$ algebra, in which there are $n-1$ holomorphic currents $W^{(k)}(z)$ with spin $k=2,3,\cdots,n$. The central charge is parametrized as
 \begin{equation} \label{eq:ccn}
  c = (n-1) - 12\, \vect{Q}\cdot\vect{Q}  = (n-1) \left( 1 -  n(n+1) Q^2 \right) \,,
\end{equation}
where $\vect{Q} = Q \vect{\rho}$ and $\vect{\rho}$ is the Weyl vector of $\mathfrak{sl}_n$ (see Appendix~\ref{app:sln_conventions}). Primary fields are characterized by their zero modes $w^{(k)}$
\begin{align}
W^{(k)}(z) \Phi(0) = \frac{w^{(k)}}{z^k} \Phi(0) + \cdots
\end{align}
and the quantum numbers $w^{(k)} = w^{(k)}_{\vect{\alpha}}$ are naturally parametrized by an $(n-1)$-dimensional vector $\vect{\alpha}$. Furthermore, they are invariant under action of the $\mathfrak{sl}_n$ Weyl group 
\begin{align}
  \forall \sigma \in \mathcal{W} \,, \qquad w^{(k)}_{\sigma \star \vect{\alpha}} = w^{(k)}_{ \vect{\alpha}} \,.
\end{align}
The dual of a charge $\vect{\alpha}$ is $\vect{\alpha}^* = - s_0 (\vect{\alpha})$, and is equivalent to $2\vect{Q} - \vect{\alpha}$. Indeed, it follows from $s_0^2 =1$ and $s_0 (\vect{\rho}) = - \vect{\rho}$ that
\begin{align}
2\vect{Q} - \vect{\alpha} = s_0 \star \left( \vect{\alpha}^* \right)
\end{align}
Of particular importance are the fully degenerate fields $\{\Phi_{b \vect\omega_1}, \Phi_{b \vect\omega_{n-1}}, \Phi_{-b^{-1} \vect\omega_1}, \Phi_{-b^{-1} \vect\omega_{n-1}}\}$. Their (chiral) fusion rules with a generic primary field is $\Phi_{\vect\alpha}$ are given by 
\begin{align}
  \Phi_{b\vect{\omega}_1} \otimes \Phi_{\vect\alpha} \rightarrow  \bigoplus_{j=1}^n \Phi_{\vect\alpha + b \vect{h}_j}, &\qquad     \Phi_{b\vect{\omega}_{n-1}} \otimes \Phi_{\vect\alpha} \rightarrow \bigoplus_{j=1}^3 \Phi_{\vect\alpha - b \vect{h}_j} \,, \\ 
    \Phi_{-b^{-1}\vect{\omega}_1} \otimes \Phi_{\vect\alpha} \rightarrow  \bigoplus_{j=1}^n \Phi_{\vect\alpha - b^{-1} \vect{h}_j},  &\qquad     \Phi_{-b^{-1}\vect{\omega}_{n-1}} \otimes \Phi_{\vect\alpha} \rightarrow \bigoplus_{j=1}^3 \Phi_{\vect\alpha + b^{-1} \vect{h}_j}  \,.
\end{align}
A particularly simple class of semi-degenerate field is given by Wyllard fields
\begin{align}
  \Phi_{\kappa \vect{\omega}_1} \quad \text{and} \quad \Phi_{\kappa \vect{\omega}_{n-1}}, \qquad \kappa \in \mathbb{C}  \,.
\end{align}
These fields have $(n-2)$ null-vectors at level one:
\begin{align}
W^{(k)}_{-1} \Phi_{\kappa \vect{\omega}_j} \propto L_{-1} \Phi_{\kappa \vect{\omega}_j}, \qquad k=3,\dots,n \,,
\end{align}
but they lack an extra null-vector to be fully degenerate. 

\section{Structure constants of scalar fields}
\label{sec:bootstrap-scalar}

So far we have given a purely chiral description of the $W_n$ theory. In order to build a consistent conformal field theory the holomorphic and anti-holomorphic sectors have to be glued appropriately. On the torus the constraint is modular invariance of the partition function, while on the sphere the constraint is crossing-symmetry of four-point functions. Here we consider the latter, and we first discuss scalar fields in the $W_n$ Toda field theory with $c \leq n-1$. Generically a field is scalar as long as its conformal spin is zero: $s = \Delta- \overline{\Delta} =0$.
In the context of an extended symmetry such as the $W_n$ theory, we take a more constraining definition, namely that all left and right quantum numbers coincide:
\begin{equation}
  \forall k \in \{2,\dots n \} \,, \quad w^{(k)} = \overline{w}^{(k)} \,. 
\end{equation}
Primary fields are labelled by their holomorphic and anti-holomorphic charge vectors as $\Phi_{\vect\alpha,\bar{\vect\alpha}}(z,\bar z)$, and scalar primary fields are those which have $\vect{\alpha} = \overline{\vect{\alpha}}$ (up to Weyl group action): we shall simply denote them as $\Phi_{\vect\alpha}(z,\bar z)$. Our purpose is to compute the structure constants of the operator algebra between scalar primary fields. These are related to the three-point correlation functions:
\begin{equation}
  C(\vect\alpha_1,\vect\alpha_2,\vect\alpha_3) = \aver{\Phi_{\vect\alpha_1}(0)\Phi_{\vect\alpha_2}(1)\Phi_{\vect\alpha_3}(\infty)} \,.
\end{equation}
When one of the vertex charges is semi-degenerate, say
\begin{equation} \label{eq:alpha3}
  \vect\alpha_3 = \kappa \vect\omega_1 
  \quad \text{or} \quad \vect\alpha_3 = \kappa \vect\omega_{n-1} \,,
  \qquad \text{with} \qquad \kappa \in \mathbb{R} \,,
\end{equation}
and $\vect\alpha_1,\vect\alpha_2$ are generic, the conformal bootstrap approach developed in~\cite{FL1,FL2} can be adapted to the imaginary Toda case. It is important to detail this calculation for the $\mathfrak{sl}_n$ Toda field theory with $c \leq n-1$. Indeed, in the case of Liouville ($n=2$) it is known that the three-point structure constants for $c\leq1$ are \emph{not} given by the analytic continuation of the ones obtained for $c \notin ] -\infty ,1]$   \cite{Zamo05,centralchargeless1}. 

In order to compute these structure constants, following \cite{FL1,FL2}, we impose crossing symmetry on the correlation function 
$$\mathcal{G}(z,\bar{z})=\aver{\Phi_{\vect\alpha_2}(\infty) \Phi_{b\vect\omega_1}(z,\bar z)  \Phi_{\vect\alpha_3}(1) \Phi_{\vect\alpha_1}(0)} \,.$$
This correlation function obeys a Fuchsian differential equation\footnote{In fact one formally treats the variables $z$ and $\bar{z}$ as being independent, and there are two differential equations : one with respect to $z$, and one with respect to $\bar{z}$.} of order $n$ (see Appendix \ref{app:hypergeometric}). 
The solutions of this differential equation form a representation of the fundamental group
\[
\rho : \pi_1 \left(    \mathbb{C}\mathbb{P}^1 \setminus \{ 0, 1, \infty \}   \right) \to \textrm{GL}(n,\mathbb{C})
 \]
The fundamental group of the punctured sphere $ \mathbb{C}\mathbb{P}^1 \setminus \{ 0, 1, \infty \} $ is generated by the loops $\gamma_0$ and $\gamma_{\infty}$  winding around $0$ and $\infty$ respectively (in the positive direction). Accordingly, we introduce two bases of solutions for the differential equation : $\{ F_i(z) , \, i =1 ,\cdots,n  \}$ with Abelian monodromies around $0$, and $\{ G_i(z) , \, i =1 ,\cdots,n  \}$ with Abelian monodromies around $\infty$. Explicit formulas in terms of hypergeometric functions can be found in Appendix \ref{app:hypergeometric}. These two bases are related through
\begin{align}
  F_i (z) = \sum_{j=1}^n M_{ij} G_j(z) \,,
\end{align}
where the matrix $M$ is given by \eqref{eq:change_basis} in Appendix \ref{app:hypergeometric}. The correlation function $\mathcal{G}(z,\bar{z})$ is built by gluing left and right conformal blocks 
\[
\label{eq:conformal_block_gluing}
\mathcal{G}(z,\bar{z}) = \sum_{i,j=1}^n X_{ij} F_i(z) F_{j} (\bar{z})
 \]
in such a way as to obtain a single-valued function of $z$, \emph{i.e.} a function with trivial monodromies. This means that for any $\lambda \in \pi_1 \left(    \mathbb{C}\mathbb{P}^1 \setminus \{ 0, 1, \infty \}   \right)$, one must have
\[
\rho(\lambda)^T X \rho(\lambda) = X
\]
It is sufficient to impose this condition for $\lambda=\gamma_0$ and $\lambda=\gamma_{\infty}$. In the generic case the blocks $F_i (z)$ have distinct monodromies, and trivial monodromy of $\mathcal{G}$ around $z=0$ is equivalent to the matrix $X$ being diagonal. Similarly, around $\infty$ one can decompose
\[
\mathcal{G}(z,\bar{z}) =  \sum_{i,j=1}^n Y_{ij} G_i(z) G_j(\bar{z}) \,,
\]
and the matrix $Y =M^{T} X M $ has to be diagonal as well. Since $X$ and $Y$ are diagonal, one has
\begin{align*}
  \sum_j X_j M_{jk} M_{jl} = Y_k \delta_{k,l}
\end{align*}
This overcomplete set of linear equations has a unique solution (up to a global prefactor)
\begin{align*}
  \frac{X_i}{X_j} = \frac{M_{j m} \left( M^{-1} \right)_{mi}}{M_{im} \left( M^{-1} \right)_{mj}}\,.
\end{align*}
provided the r.h.s. does not depend on $m$. For scalar fields this is indeed the case, since
\begin{align*}
  \frac{M_{j m} \left( M^{-1} \right)_{mi}}{M_{im} \left( M^{-1} \right)_{mj}}= \frac{\prod_{k \neq i} \gamma(A_k - A_i)}{ \prod_{k \neq j}\gamma(A_k- A_j)} \prod_{k=1}^n \frac{\gamma(A_i+B_k)}{\gamma(A_j+B_k)} \,,
\end{align*}
where $\gamma(x) = \Gamma(x)/\Gamma(1-x)$,
\begin{equation}
  A_i = \Delta_{\vect\alpha_1 + b \vect{h}_i} - \Delta_{\vect\alpha_1} + b\mu \,, \qquad
  B_i = \Delta_{\vect\alpha_2 + b \vect{h}_i} - \Delta_{\vect\alpha_2} + b\mu \,,
\end{equation}
and
\begin{equation}
  \label{eq:mu}
  2 \mu = 
  \begin{cases}
    (1/b - b)-\frac{\kappa}{n} \quad & \text{ if }  \ \vect\alpha_3=\kappa \vect\omega_1 \,, \\
    \frac{\kappa}{n} \quad  & \text{ if }  \  \vect\alpha_3=\kappa \vect\omega_{n-1} \,.
  \end{cases}
\end{equation}
The coefficients $X_i$ are related to the structure constants as follows
\begin{align*}
  \frac{X_i}{X_j} = \frac{C(\vect{\alpha}_1,b \vect{\omega}_1, 2 \vect{Q} - \vect{\alpha}_1 - b \vect{h}_i )}{C(\vect{\alpha}_1,b \vect{\omega}_1, 2 \vect{Q} - \vect{\alpha}_1 - b \vect{h}_j )}\frac{C(\vect{\alpha}_1+b \vect{h}_i,\vect{\alpha}_2,\vect{\alpha}_3)}{C(\vect{\alpha}_1+b \vect{h}_j,\vect{\alpha}_2,\vect{\alpha}_3)} \,,
\end{align*}
and crossing symmetry boils down to
\begin{align} \label{eq:shift-scalar}
  \frac{C(\vect{\alpha}_1+b \vect{h}_i,\vect{\alpha}_2,\vect{\alpha}_3)}{C(\vect{\alpha}_1+b \vect{h}_j,\vect{\alpha}_2,\vect{\alpha}_3)}=
  \frac{K_i(\vect\alpha_1)}{K_j(\vect\alpha_1)}
  \times \prod_{k=1}^n \frac{\gamma(A_i+B_k)}{\gamma(A_j+B_k)} \,,
\end{align}
where $K_i(\vect\alpha_1)$ and $K_j(\vect\alpha_1)$ are some normalising factors.
Repeating the same steps with the fully degenerate field $\Phi_{-b^{-1} \vect{\omega}_1}$ yields the same relation with $b \to -b^{-1}$:
\begin{align} \label{eq:shift-scalar2}
  \frac{C(\vect{\alpha}_1- \vect{h}_i/b,\vect{\alpha}_2,\vect{\alpha}_3)}
       {C(\vect{\alpha}_1- \vect{h}_j/b,\vect{\alpha}_2,\vect{\alpha}_3)}=
  \frac{\wh K_i(\vect\alpha_1)}{\wh K_j(\vect\alpha_1)}
  \times \prod_{k=1}^n \frac{\gamma(C_i+D_k)}{\gamma(C_j+D_k)} \,,
\end{align}
where $\wh K_i(\vect\alpha_1)$ and $\wh K_j(\vect\alpha_1)$ are some normalising factors, and
\begin{equation}
  C_i = \Delta_{\vect\alpha_1 -\vect{h}_i/b} - \Delta_{\vect\alpha_1} -\mu/b \,, \qquad
  D_i = \Delta_{\vect\alpha_2 -\vect{h}_i/b} - \Delta_{\vect\alpha_2} -\mu/b \,.
\end{equation}

If $b$ is real and $b^2$ is irrational then these two relations determine uniquely the three point function up to a $\kappa$ dependent multiplicative factor. It can be expressed in terms of the $\Upsilon_b$ function, whose definition and main properties we recall in appendix \ref{app:Upsilon}:
\begin{align}
  \label{eq:C-vertex}
  C(\vect\alpha_1,\vect\alpha_2,\vect\alpha_3) = M(\kappa) \times 
  \frac{\prod_{k,\ell=1}^n \Upsilon_{b} \left[b + (\vect{\alpha}_1- \vect{Q})\cdot \vect{h}_k +(\vect{\alpha}_2- \vect{Q})\cdot \vect{h}_\ell + 2 \mu \right]}
       {\sqrt{\prod_{i=1}^2 \prod_{\vect{e}>0} \Upsilon_{b}\left[b + (\vect{\alpha}_i-\vect{Q})\cdot \vect{e}  \right] \Upsilon_{b}\left[b - (\vect{\alpha}_i-\vect{Q})\cdot \vect{e}  \right]}} \,,
\end{align}
where $\vect\alpha_3$ must be semi-degenerate and $\mu$ is given in \eqref{eq:mu}. In the denominator the product is over all positive roots of $\mathfrak{sl}_n$.
The full bootstrap argument above is only valid when $\vect\alpha_3$ is semidegenerate, \emph{and} $\vect\alpha_1, \vect\alpha_2$ are nondegenerate, i.e. they are not of the form \eqref{eq:alpha3}. Indeed, if for instance $\vect\alpha_1$ is semi-degenerate (say $\vect\alpha_1=\lambda\vect\omega_1$), then in the channel $z \to 0$ only two conformal blocks are allowed for $\mathcal{G}$, due to the fusion rule
$$\Phi_{b \vect{\omega}_1} \otimes \Phi_{\lambda \vect{\omega}_1} \rightarrow \Phi_{\lambda \vect{\omega}_1 + b\vect{h}_1} \oplus \Phi_{\lambda \vect{\omega}_1 + b\vect{h}_2} \,,$$
and then the above derivation of structure constants is no longer valid. An exception is when $\vect\alpha_2=0$: in this case one considers the OPE of a semidegenerate field $\Phi_{\vect\alpha_3}$ with the identity.

Note that it is not necessary to keep track of the normalising factors in (\ref{eq:shift-scalar}--\ref{eq:shift-scalar2}) to derive \eqref{eq:C-vertex}. Any function of the form
\begin{equation*}
  C(\vect\alpha_1,\vect\alpha_2,\vect\alpha_3) = \frac{\prod_{k,\ell=1}^n \Upsilon_{b} \left[b + (\vect{\alpha}_1- \vect{Q})\cdot \vect{h}_k +(\vect{\alpha}_2- \vect{Q})\cdot \vect{h}_\ell + 2 \mu \right]}
       {\phi(\vect\alpha_1) \phi(\vect\alpha_2)}
\end{equation*}
is a solution of (\ref{eq:shift-scalar}--\ref{eq:shift-scalar2}) with some functions $\{K_i(\vect\alpha)\}$ and $\{\wh K_i(\vect\alpha)\}$ given in terms of $\phi(\vect\alpha)$, and the precise form of $\phi(\vect\alpha)$ is completely determined by imposing:
\begin{align}
  C(\vect{\alpha}_1,\vect{\alpha}_1^*,0) & = 1 \\
  C(\vect{\alpha}_1,\vect{\alpha}_2, \kappa \vect{\omega}_1) & =
  C(\vect{\alpha}^*_1,\vect{\alpha}_2^*, \kappa \vect{\omega}_{n-1}) \,.
\end{align}
The factor $M(\kappa)$ can be then be found by demanding that $C(\kappa \vect{\omega}_1,0,\kappa\vect{\omega}_{n-1})=1$, which yields:
\begin{equation} \label{eq:M}
  M(\kappa) = \frac{1}{\Upsilon_b(b)^n} \sqrt{\frac{\Upsilon_b (b) \Upsilon_b (b + nQ)}{\Upsilon_b (b+ \kappa) \Upsilon_b (b - \kappa + nQ)}} \,.
\end{equation}
The three point function\eqref{eq:C-vertex} enjoys the following properties:
\begin{align}
  C(\vect{\alpha}_1,\vect{\alpha}_2, \vect\alpha_3) & =
  C(\vect{\alpha}_2,\vect{\alpha}_1,  \vect\alpha_3) \,, \\
  \forall \sigma,\sigma' \in \mathfrak{S}_n^2 \,, \qquad
  C(\vect{\alpha}_1,\vect{\alpha}_2,\kappa \vect\alpha_3) & = C(\sigma\star \vect{\alpha}_1,\sigma' \star\vect{\alpha}_2,\kappa \vect\alpha_3) \,.
\end{align}

Finally, to compare, the three-point function found by Fateev and Litvinov~\cite{FL1,FL2} in the case of real Toda with central charge $c=(n-1)+12\,\vect{\widehat{Q}}^2$ is of the form:
\begin{equation} \label{eq:DOZZ_general}
  C_{\rm FL}(\vect\alpha_1, \vect\alpha_2, \kappa\vect\omega_{n-1}) =
  \frac{A(\vect\alpha_1) A(\vect\alpha_2) B(\kappa)}
       {\prod_{i,j} \Upsilon_{\hat{b}}\left(\nicefrac{\kappa}{3} + (\vect{\alpha}_1 - \vect{\widehat{Q}})\cdot \vect{h}_i + (\vect{\alpha}_2 - \vect{\widehat{Q}})\cdot \vect h_j\right)}
\end{equation}
where $\vect{\hat{Q}} = \left( \hat{b} + \nicefrac{1}{\hat{b}} \right) \vect \rho$, and $A(\vect\alpha)$ and $B(\kappa)$ are some normalising factors. We see that, like for the Liouville theory, the structure constants of the imaginary Toda theory are \emph{not} the analytic continuation of the real Toda ones.

\section{Non-scalar fields in the imaginary Toda field theory}

In the previous section we considered scalar fields parametrized by the same  charges $\vect{\alpha} = \bar{\vect{\alpha}}$ in the holomorphic and anti-holomorphic sectors, in a $W_n$-conformal field theory with a generic central charge $c \leq n-1$ (\emph{i.e.} $b^2$ non rational). Scalar fields are mutually local, and this leads to monodromy invariant correlation function. But it is also possible for a physical correlation function to acquire a non-trivial phase as a field winds around another.  For instance this is typically the case for spin and disorder operators in the $\mathbb{Z}_{n}$ parafermion model, in which the phase is then a $n^{\textrm{th}}$ root of unity. 

This motivates an investigation of non-scalar primary fields $\Phi_{\vect{\alpha}, \bar{\vect{\alpha}}}$ parametrised by two vector charges $\vect{\alpha}$ and $\bar{\vect{\alpha}}$.

\subsection{Consistency conditions on OPEs}
\label{sec:consistency}

The fully degenerate fields $\{\Phi_{b \vect{\omega}_1}, \Phi_{b \vect{\omega}_{n-1}}, \Phi_{-b^{-1} \vect{\omega}_1}, \Phi_{-b^{-1} \vect{\omega}_{n-1}}\}$ are assumed to be part of the spectrum of the theory.  By demanding a well-defined monodromy between $\Phi_{\vect{\alpha}, \bar{\vect{\alpha}}}$ and these fully degenerate fields, some constraints are obtained on the possible values of $\vect{\alpha}$ and $\bar{\vect{\alpha}}$. Consider for instance the OPE between $\Phi_{b \vect{\omega}_1}$ and $\Phi_{\vect{\alpha}, \bar{\vect{\alpha}}}$. Since we know the (chiral) fusion rules of $\Phi_{b \vect{\omega}_1}$ with any field, this OPE has to be of the form
\[
\Phi_{b\vect{\omega}_1}(z,\bar{z}) \,\Phi_{\vect\alpha, \bar{\vect\alpha}}(0)
= \sum_{i,j=1}^n  \mathcal{C}_{ij}
\, z^{\Delta_{\vect{\alpha} + b\vect{h}_i}  - \Delta_{b \vect\omega_1} - \Delta_{\vect\alpha}}
\, \bar{z}^{\Delta_{\vect{\bar\alpha}+ b\vect{h}_j}  - \Delta_{b \vect\omega_1} - \Delta_{\vect{\bar\alpha}} }
\, \Phi_{\vect{\alpha} + b \vect{h}_{i}, \bar{\vect{\alpha}} + b \vect{h}_j}(0)  + \cdots
\]
as long as the field $\Phi_{\vect\alpha, \bar{\vect\alpha}}$ is not semi- or fully-degenerate in the sense of \secref{degenerate-rep}.
We impose that every term in the right-hand side has the same monodromy $e^{2i\pi\eta}$ when $z$ goes around zero, so that:

\begin{align}
  \Phi_{b\vect{\omega}_1}\left(e^{2i\pi }z,e^{-2i\pi } \bar{z} \right)\Phi_{\vect\alpha, \bar{\vect\alpha}}(0) = e^{2i\pi \eta} \, \Phi_{b\vect{\omega}_1}\left(z,\bar{z} \right)\Phi_{\vect\alpha, \bar{\vect\alpha}}(0) \,.
\end{align}
The monodromy exponent for the term $(i,j)$ in the above sum is
$$\eta_{ij}=(\Delta_{\vect\alpha+b\vect h_i}-\Delta_{\vect{\bar\alpha}+b\vect h_j})-(\Delta_{\vect\alpha}-\Delta_{\vect{\bar\alpha}}) \,.$$
Let us consider the case when all the exponents $\eta_{ij}$ are distinct modulo one, {\it i.e.} when $(\eta_{ik}-\eta_{jk})$ and $(\eta_{ki}-\eta_{kj})$ are \emph{not} integers if $i \neq j$, which happens if and only if:
\begin{equation} \label{eq:alpha-generic}
  \forall i \neq j \,, \qquad
  (\vect\alpha-\vect Q) \cdot (\vect h_i-\vect h_j) \notin \mathbb{Z}/b \,
  \qquad \text{and} \qquad
  (\vect{\bar\alpha}-\vect Q) \cdot (\vect h_i-\vect h_j) \notin \mathbb{Z}/b \,.
\end{equation}
We shall refer to this situation by saying that both $\vect\alpha$ and $\vect{\bar\alpha}$ are \textit{generic}. In this case, the coefficient matrix must be of the form:
\[
\mathcal{C}_{ij} = \delta_{i,\tau(j)} \mathcal{C}_{j} \,,
\] 
where $\tau \in  \mathfrak{S}_n$ is a permutation. This permutation encodes the fusion rules 
\begin{equation} \label{eq:fusion}
\Phi_{b\vect{\omega}_1} \times \Phi_{\vect\alpha, \bar{\vect\alpha}} = \sum_{j=1}^n \Phi_{\vect{\alpha} + b \vect{h}_{\tau(j)}, \bar{\vect{\alpha}}+ b \vect{h}_j} \,. 
\end{equation}
Note that, using~\eqref{eq:relabel}, we are free to relabel $\vect\alpha \to \vect{\alpha'} =\mu \star \vect\alpha$, and $\vect{\bar\alpha} \to \vect{\bar\alpha'} =\bar\mu \star \vect{\bar\alpha}$, to get:
\begin{align}
  \Phi_{b\vect{\omega}_1} \times \Phi_{\vect{\alpha'}, \vect{\bar\alpha'}}
  &\equiv \Phi_{b\vect{\omega}_1} \times \Phi_{\vect\alpha, \vect{\bar\alpha}}
  = \sum_{j=1}^n \Phi_{\vect{\alpha} + b \vect{h}_{\tau(j)}, \bar{\vect{\alpha}}+ b \vect{h}_j} \nn \\
  & \equiv \sum_{j=1}^n \Phi_{\mu\star(\vect{\alpha} + b \vect{h}_{\tau(j)}), \bar\mu \star(\bar{\vect{\alpha}}+ b \vect{h}_j)}
  = \sum_{j=1}^n \Phi_{\vect{\alpha'} + b \vect{h}_{\mu\tau(j)}, \vect{\bar\alpha'}+ b \vect{h}_{\bar\mu(j)}} \nn \\
  &= \sum_{k=1}^n \Phi_{\vect{\alpha'} + b \vect{h}_{\mu\tau\bar\mu^{-1}(k)}, \vect{\bar\alpha'}+ b \vect{h}_k} \,.
  \label{eq:conjug}
\end{align}

In the OPE, which we now write as
\[
\Phi_{b\vect{\omega}_1}(z,\bar{z})\Phi_{\vect\alpha, \bar{\vect\alpha}}(0) = \sum_{j=1}^n  \mathcal{C}_j \,
z^{\Delta_{\vect\alpha + b\vect{h}_{\tau(j)}}  - \Delta_{b \vect\omega_1} - \Delta_{\vect\alpha}} \,
\bar{z}^{\Delta_{\vect{\bar\alpha}+ b\vect{h}_j}  - \Delta_{b \vect\omega_1} - \Delta_{\vect{\bar\alpha}}} \,
\Phi_{\vect{\alpha} + b \vect{h}_{\tau(j)}, \bar{\vect{\alpha}} + b \vect{h}_j}  + \cdots
\]
the monodromy exponent for the $j^{\textrm{th}}$ term in the sum is
\begin{align*}
  \eta_j:=\eta_{\tau(j),j}=(\Delta_{\vect\alpha+b\vect h_{\tau(j)}}-\Delta_{\vect{\bar\alpha}+b\vect h_j})-(\Delta_{\vect\alpha}-\Delta_{\vect{\bar\alpha}}) = b \vect{h}_j \cdot (\tau^{-1} \star \vect\alpha - \bar{\vect\alpha}) \,.
\end{align*}

Since the vectors $(\vect h_i-\vect h_j)$ generate the lattice root lattice $\mathcal{R}$ (see Appendix~\ref{app:sln_conventions}), the condition $\eta_j - \eta_k \in \mathbb{Z}$ for all $j$ and $k$ boils down to:
\begin{equation}
  \label{eq:condition_alpha_a1}
  \tau^{-1} \star \vect{\alpha} - \vect{\bar{\alpha}} \in  \mathcal{R}^*/b
  \quad \Leftrightarrow \quad \vect\alpha - \tau \star \vect{\bar\alpha} \in  \mathcal{R}^*/b \,,
\end{equation}
where $\mathcal{R}^*$ is the weight lattice. An interesting consequence is that 
\begin{align*}
  \eta \in \frac{\mathbb{Z}}{n} \,,
\end{align*}
\emph{i.e.} the overall monodromy around $z=0$ of the OPE $\Phi_{b\vect{\omega}_1}(z,\bar{z})\Phi_{\vect\alpha, \bar{\vect\alpha}}(0) $ can only be a $n^{\textrm{th}}$ root of unity. 
One can repeat the above arguments with the fully-degenerate field $\Phi_{b\vect\omega_1}$ replaced by its dual $\Phi_{b \vect{\omega}_{n-1}}$, by simply noting that the monodromy exponents $\eta_j$ get an overall factor of $(-1)$. Hence, the OPE coefficients $\mathcal{C}_{ij}$ are determined by the same permutation $\tau$, and the fusion rules can be written:
\begin{align*}
  \Phi_{b\vect{\omega}_{n-1}} \times \Phi_{\vect\alpha, \bar{\vect\alpha}} = \sum_{j=1}^n \Phi_{\vect{\alpha} - b \vect{h}_{\tau(j)}, \bar{\vect{\alpha}}- b \vect{h}_j} \,.
\end{align*}

Let us now examine the fusion of our $\Phi_{\vect\alpha, \vect{\bar\alpha}}$ with the fields $\Phi_{-\vect{\omega}_1/b}$ and $\Phi_{-\vect{\omega}_{n-1}/b}$. We want to consider the generic situation, as in~\eqref{eq:alpha-generic} :
\begin{equation} \label{eq:alpha-generic2}
  \forall i \neq j \,, \qquad
  (\vect\alpha-\vect Q) \cdot (\vect h_i-\vect h_j) \notin b\mathbb{Z} \,
  \qquad \text{and} \qquad
  (\vect{\bar\alpha}-\vect Q) \cdot (\vect h_i-\vect h_j) \notin b\mathbb{Z} \,.
\end{equation}
The same line of reasoning as above yields:
\begin{equation}
  \label{eq:condition_alpha_b}
  \sigma^{-1} \star \vect{\alpha} -  \vect{\bar{\alpha}} \in  b\mathcal{R}^*
  \quad \Leftrightarrow \quad  \vect\alpha - \sigma \star \vect{\bar\alpha} \in b\mathcal{R}^*\,,
\end{equation}
corresponding to the fusion rules 
\begin{align}
  \Phi_{-\vect{\omega}_1/b} \times \Phi_{\vect\alpha, \bar{\vect\alpha}}  & = \sum_{j=1}^n \Phi_{\vect{\alpha} - \vect{h}_{ \sigma (j)/b}, \bar{\vect{\alpha}} - \vect{h}_j/b} \,, \\
  \Phi_{-\vect{\omega}_{n-1}/b} \times \Phi_{\vect\alpha, \bar{\vect\alpha}} & = \sum_{j=1}^n \Phi_{\vect{\alpha} + \vect{h}_{ \sigma(j)/b}, \bar{\vect{\alpha}} + \vect{h}_j/b} \,,
\end{align}
with some permutation $\sigma \in \mathfrak{S}_n$, possibly different from $\tau$.
Therefore, it appears that a generic primary field is labelled by two charges $(\vect{\alpha},\bar{\vect{\alpha}})$ and two permutations $(\tau,\sigma)$.
But we still have to discuss the effect of charge reparametrisation by the Weyl group \eqref{eq:relabel}. As seen in~\eqref{eq:conjug}, if we relabel $(\vect\alpha, \vect{\bar\alpha}) \to (\vect{\alpha'}, \vect{\bar\alpha'}) = (\mu\star \vect\alpha, \bar\mu\star \vect{\bar\alpha})$, the permutations $\tau$ and $\sigma$ are changed to:
\begin{equation}
  \tau \to \mu \tau \bar\mu^{-1} \,,
  \qquad \sigma \to \mu \sigma \bar\mu^{-1} \,.
\end{equation}
Taking a generic permutation $\mu$ and setting $\bar\mu=\mu\tau$, we get
\begin{equation}
  \tau \to \id \,,
  \qquad \sigma \to \mu \sigma' \mu^{-1} \,,
\end{equation}
where $\sigma'=\sigma\tau^{-1}$. Hence, without loss of generality, $\tau$ can always be set to the identity, whereas $\sigma$ is defined modulo conjugation by any permutation $\mu$.

Note that the fusion of a generic field $\Phi_{\vect\alpha,\vect{\bar\alpha}}$ [i.e. a field satisfying \eqref{eq:alpha-generic} and \eqref{eq:alpha-generic2}] with any of the degenerate fields $\Phi_{b\omega_1}, \Phi_{b\omega_{n-1}}, \Phi_{-\omega_1/b}, \Phi_{-\omega_{n-1}/b}$ may produce non-generic fields. However, in the typical case when the terms in the right-hand side of~\eqref{eq:fusion} are generic, one may ask what permutation they correspond to. This is easy to see from the constraints~\eqref{eq:condition_alpha_a1} and \eqref{eq:condition_alpha_b} on vertex charges. We set $\tau$ to the identity, so that \eqref{eq:condition_alpha_a1} becomes $\vect\alpha-\vect{\bar\alpha} \in \mathcal{R}/b$. This condition is obviously satisfied by $(\vect\alpha+b\vect{h}_j,\vect{\bar\alpha}+b\vect{h}_j)$. The second condition~\eqref{eq:condition_alpha_b} for this pair of charges reads
$$ (\vect\alpha - \sigma \star \vect{\bar\alpha}) + b(\vect{h}_j-\vect{h}_{\sigma(j)}) \in b\mathcal{R}^* \,, $$
which is also satisfied, because every $\vect{h}_k$ belongs to $\mathcal{R}^*$. Similar arguments hold for the other fusions considered above, and for the dual field $\Phi_{2\vect{Q}-\vect\alpha,2\vect{Q}-\vect{\bar\alpha}}$.

\subsection{Characterisation of generic operators}
\label{sec:classif}

Overall, from the arguments of \secref{consistency}, we get the following characterisation of generic primary fields:
\begin{itemize}
\item A primary field is labelled by a pair of vertex charges $(\vect\alpha,\vect{\bar\alpha})$, and a  permutation $\sigma \in \mathfrak{S}_n$. We denote it as $\Phi_{\vect\alpha,\vect{\bar\alpha}}^{(\sigma)}$, and  the vertex charges must satisfy:
\begin{align}
  \label{eq:constraints}
  \vect{\alpha} -  \vect{\bar{\alpha}} \in \mathcal{R}^*/b \,,
  \qquad \text{and} \qquad \vect{\alpha} - \sigma \star \vect{\bar{\alpha}} \in  b \mathcal{R}^* \,.
\end{align}

Note that these conditions, as well as the quantum numbers $w^{(k)}_{\vect\alpha}, w^{(k)}_{\vect{\bar\alpha}}$, are invariant under reparameterisation $(\vect\alpha,\vect{\bar\alpha},\sigma) \to (\mu\star\vect\alpha,\mu\star\vect{\bar\alpha},\mu\sigma\mu^{-1})$. Hence, the behaviour of $\Phi_{\vect\alpha,\vect{\bar\alpha}}^{(\sigma)}$ under fusion is really determined by the \emph{conjugacy class} of $\sigma$.

\item The fusion rules with the fully degenerate fields are:
  \begin{align}
    \Phi_{b\vect{\omega}_1} \times \Phi^{(\sigma)}_{\vect\alpha, \bar{\vect\alpha}}  & = \sum_{j=1}^n \Phi^{(\sigma)}_{\vect{\alpha} + b \vect{h}_{j}, \bar{\vect{\alpha}}+ b \vect{h}_j} \,,
    &&& \Phi_{b\vect{\omega}_{n-1}} \times \Phi^{(\sigma)}_{\vect\alpha, \bar{\vect\alpha}}  & = \sum_{j=1}^n \Phi^{(\sigma)}_{\vect{\alpha} - b \vect{h}_{j}, \bar{\vect{\alpha}}- b \vect{h}_j} \,, \label{eq:fusion1} \\
    \Phi_{-\vect{\omega}_1/b} \times \Phi^{(\sigma)}_{\vect\alpha, \bar{\vect\alpha}}  & = \sum_{j=1}^n \Phi^{(\sigma)}_{\vect{\alpha} - \vect{h}_{ \sigma (j)}/b, \bar{\vect{\alpha}} - \vect{h}_j/b} \,,
    &&& \Phi_{-\vect{\omega}_{n-1}/b} \times \Phi^{(\sigma)}_{\vect\alpha, \bar{\vect\alpha}}  & = \sum_{j=1}^n \Phi^{(\sigma)}_{\vect{\alpha} + \vect{h}_{ \sigma (j)}/b, \bar{\vect{\alpha}} + \vect{h}_j/b} \,. \label{eq:fusion2} 
  \end{align}
  The monodromy exponents (defined up to the addition of an integer) of the corresponding OPEs are :
  \begin{equation} \label{eq:eta}
    \eta(\vect\alpha,\vect{\bar\alpha}) = b \vect{h}_1 \cdot (\vect\alpha-\vect{\bar\alpha}) \,,
    \qquad \wh\eta(\vect\alpha,\vect{\bar\alpha}) = -\frac{1}{b} \vect{h}_1 \cdot (\vect\alpha-\sigma \star \vect{\bar\alpha}) \,,
  \end{equation}
  for the fusion of $\Phi^{(\sigma)}_{\vect\alpha, \bar{\vect\alpha}}$ with $\Phi_{b\vect{\omega}_1}$ and $\Phi_{-\vect{\omega}_1/b}$, respectively. These exponents belong to $\mathbb Z/n$, and the monodromy factors $e^{2i\pi\eta(\vect\alpha,\vect{\bar\alpha})}$ and $e^{2i\pi\wh\eta(\vect\alpha,\vect{\bar\alpha})}$ can be considered as two $\mathbb{Z}_n$ charges associated to the field $\Phi^{(\sigma)}_{\vect\alpha, \bar{\vect\alpha}}$.

\item The dual of the field $\Phi^{(\sigma)}_{\vect\alpha, \bar{\vect\alpha}}$ is : $\left(\Phi^{(\sigma)}_{\vect\alpha, \bar{\vect\alpha}} \right)^* = \Phi^{(\sigma)}_{2\vect{Q} - \vect\alpha, 2\vect{Q} -  \bar{\vect\alpha}}$.

\item The particular case $\sigma= \id$ corresponds to \emph{scalar} primary fields. Indeed, as long as $b^2$ is not rational, the conditions \eqref{eq:constraints} yield $\vect\alpha = \vect{\bar\alpha}$.
  
\end{itemize}

Note that this characterisation applies only to \emph{generic}, nondegenerate values of the vertex charges $(\vect\alpha,\vect{\bar\alpha})$, i.e. nondegenerate charges satisfying \eqref{eq:alpha-generic} and \eqref{eq:alpha-generic2}.

\subsection{The case of semi-degenerate operators}

For a semi-degenerate field, which we shall denote $\wt\Phi_{\vect\alpha,\vect{\bar\alpha}}$, we have to use the fusion rules~\eqref{eq:fusion_rules_degenerate_semi_degenerate_sln}. For instance, for $(\vect \alpha, \vect{\bar\alpha}) = (\kappa \vect\omega_1 , \bar\kappa \vect\omega_1)$, one has the chiral fusion rule:
\[
\Phi_{b\vect\omega_1} \otimes \Phi_{\kappa\vect\omega_1} \to
\Phi_{\kappa\vect\omega_1 + b \vect{h}_1} + (n-1) \Phi_{\kappa\vect\omega_1 + b \vect{h}_2} \,,
\]
where the coefficient on the second term means that there are $(n-1)$ independent conformal blocks corresponding to this internal field in the fusion channel $z \to 1$ for any four-point function of the form~\eqref{eq:non-diagonal_correlation_function}. For generic values of $\kappa$, the monodromy exponents corresponding to the first term and the next $(n-1)$ do not differ by an integer. Hence, in order to get a well-defined monodromy for the four-point function, one has to select the fusion rule:
\[
\Phi_{b\vect\omega_1} \otimes \wt\Phi_{\kappa\vect\omega_1,\bar\kappa\vect\omega_1} \to
\Phi_{\kappa\vect\omega_1 + b \vect{h}_1,\bar\kappa\vect\omega_1 + b \vect{h}_1} + (n-1) \Phi_{\kappa\vect\omega_1 + b \vect{h}_2,\bar\kappa\vect\omega_1 + b \vect{h}_2} \,.
\]
From there, using a similar computation as for the case of generic non-scalar operators (see previous section), one gets the constraint:
\begin{equation}
  \kappa - \bar\kappa \in \mathbb{Z}/b \,.
\end{equation}

\subsection{Structure constants}

\subsubsection{Shift equation from the null descendant of $\Phi_{b\vect\omega_1}$}

In order to compute the structure constants, we turn to the four point function 
\begin{align}
  \label{eq:non-diagonal_correlation_function}
  &\mathcal{G}(z,\bar{z}) =  \aver{\Phi^{(\sigma_2)}_{\vect{\alpha}_2^* , \vect{\bar\alpha}_2^*}(\infty) \Phi_{b\vect{\omega}_1}(z,\bar{z}) \wt\Phi_{\vect \alpha_3, \vect{\bar\alpha}_3}(1) \Phi^{(\sigma_1)}_{\vect{\alpha}_1 , \vect{\bar\alpha}_1}(0)} \,,
\end{align}
where
$$(\vect \alpha_3, \vect{\bar\alpha}_3) =(\kappa \vect\omega_1 , \bar\kappa \vect\omega_1)
\quad \text{or} \quad (\vect \alpha_3, \vect{\bar\alpha}_3) =(\kappa \vect\omega_{n-1} , \bar\kappa \vect\omega_{n-1}) \,, \qquad \text{with} \qquad (\kappa,\bar\kappa) \in \mathbb{R}^2 \,.$$
For this correlation function to be non-trivial, one needs to impose the constraint on monodromy exponents~\eqref{eq:eta}:
\begin{equation} \label{eq:total-mono}
  e^{2i\pi[\eta(\vect\alpha_1, \vect{\bar\alpha}_1) + \eta(\vect\alpha_2, \vect{\bar\alpha}_2) + \eta(\vect\alpha_3, \vect{\bar\alpha}_3)]}= 1 \,,
\end{equation}
which can be viewed as a $\mathbb{Z}_n$ charge neutrality condition on $\cal G$.

The correlation function \eqref{eq:non-diagonal_correlation_function} is built by gluing the left and right conformal blocks in such a way as to ensure well-defined global monodromies : 
\[
\mathcal{G}(z,\bar{z}) = \sum_{i,j=1}^n X_{ij} F_i(z) \bar{F}_{j} (\bar{z}) = \sum_{i,j=1}^n Y_{ij} G_i(z) \bar{G}_{j} (\bar{z}) \,,
\]
where the conformal blocks are the same as in the scalar case, and are given in appendix \ref{app:hypergeometric}.  For the reader's convenience, we recall that the  holomorphic blocks $F_i$ and $G_i$ are expressed in terms of 
\begin{equation}
  \begin{split}
    A_i &= \Delta_{\vect\alpha_1 + b \vect{h}_i} - \Delta_{\vect\alpha_1} + b\mu \,,
    \qquad \overline{A}_i = \Delta_{\vect{\bar\alpha}_1 + b \vect{h}_i} - \Delta_{\vect{\bar\alpha}_1} + b\overline{\mu} \,, \\
    B_i &= \Delta_{\vect\alpha_2 + b \vect{h}_i} - \Delta_{\vect\alpha_2} + b\mu   \,,
    \qquad \overline{B}_i = \Delta_{\vect{\bar\alpha}_2 + b \vect{h}_i} - \Delta_{\vect{\bar\alpha}_2} + b\overline{\mu} \,,
  \end{split}
\end{equation}
where
\begin{equation}
  (2 \mu,2\bar\mu) = 
  \begin{cases}
    \left[(1/b - b)-\frac{\kappa}{n}, (1/b - b)-\frac{\bar\kappa}{n} \right]  \quad & \text{ if }  \ (\vect \alpha_3, \vect{\bar\alpha}_3) = (\kappa \vect\omega_1 , \bar\kappa \vect\omega_1) \,, \\
    (\frac{\kappa}{n},\frac{\bar\kappa}{n})  \quad  & \text{ if }  \  (\vect \alpha_3, \vect{\bar\alpha}_3) = (\kappa \vect\omega_{n-1} , \bar\kappa \vect\omega_{n-1}) \,.
  \end{cases}
\end{equation}

From the fusion rules of non-scalar fields with $\Phi_{b \vect{\omega}_1}$ one must have $X_{ij} = X_i \delta_{i,j}$ and $Y_{i,j} = Y_i \delta_{ij}$.
At this point it is interesting to compute the differences:
\begin{equation} \label{eq:A-Ab}
  A_i-\overline{A}_i = b \vect{h}_i \cdot (\vect\alpha_1-\vect{\bar\alpha}_1) + b(\mu-\bar\mu) \,,
  \qquad B_i-\overline{B}_i = b \vect{h}_i \cdot (\vect\alpha_2-\vect{\bar\alpha}_2) + b(\mu-\bar\mu) \,.
\end{equation}
Since $(\vect\alpha_k-\vect{\bar\alpha}_k) \in \mathcal{R}^*/b$, these quantities are independent of $i$ (up to an integer), and, due to \eqref{eq:total-mono}, one has:
\begin{equation} \label{eq:total-mono2}
  (A_i-\overline{A}_i) + (B_j-\overline{B}_j) \in \mathbb{Z} \,.
\end{equation}
Now if we apply the change of bases between $0$ and $\infty$ to $\sum_j Y_j |J_{j}(z)|^2$, we find
$$Y = M^t \, X \, \overline{M} \,,$$
where the matrices $M = M(\{A_i\},\{B_j\})$ and $\overline{M} = M(\{\overline{A}_i\},\{\overline{B}_j\})$ are given by \eqref{eq:change_basis}. Since $X$ and $Y$ are diagonal, one has for all $k$ and $\ell$:
\begin{equation} \label{eq:MXM=Y}
  \sum_j M_{jk} X_j \overline{M}_{j\ell} = Y_k \delta_{k\ell} \,.
\end{equation}
The subsystem of equations corresponding to fixed $\ell$ and $k=1, \dots, n$ yields:
\begin{equation}
  \label{eq:non_diagonal_bootstrap}
  X_j \propto \frac{\left(M^{-1}\right)_{\ell j}}{\overline{M}_{j\ell}}
  \qquad \Rightarrow\qquad \forall (i,j) \,, \quad \frac{X_i}{X_j} = \frac{\overline{M}_{j \ell} \left( M^{-1} \right)_{\ell i}}{\overline{M}_{i\ell} \left( M^{-1} \right)_{\ell j}} \,.
\end{equation}
For a non-trivial solution to exist, the ratio $X_i/X_j$ should be independent of $\ell$:
\begin{equation*}
  \forall (i,j,\ell,m) \,, \qquad
  \frac{\overline{M}_{j \ell} \left( M^{-1} \right)_{\ell i}}{\overline{M}_{i\ell} \left( M^{-1} \right)_{\ell j}}
  = \frac{\overline{M}_{j m} \left( M^{-1} \right)_{mi}}{\overline{M}_{im} \left( M^{-1} \right)_{mj}}\,,
\end{equation*}
which boils down to the consistency condition:
\begin{equation} \label{eq:bootstrap-consist}
  \frac{\sin \pi (\overline{A}_i + \overline{B}_\ell) \sin \pi (A_j+B_\ell)}
       {\sin \pi (A_i+B_\ell) \sin \pi (\overline{A}_j + \overline{B}_\ell)}
       = \frac{\sin \pi (\overline{A}_i + \overline{B}_m) \sin \pi (A_j+B_m)}
       { \sin \pi (A_i+B_m) \sin \pi (\overline{A}_j + \overline{B}_m)} \,.
\end{equation}
This is the $\mathfrak{sl}_n$ generalisation of the constraint obtained in \cite{EI15} in the $\mathfrak{sl}_2$ case. The property~\eqref{eq:total-mono2} deriving from the single constraint~\eqref{eq:total-mono} is actually a sufficient condition for \eqref{eq:bootstrap-consist} to be satisfied.

Let us now turn to \eqref{eq:non_diagonal_bootstrap}. This translates into the following shift equation:
\begin{align}
  &\frac{C(\Phi^{(\sigma_1)}_{\vect\alpha_1,\vect{\bar\alpha}_1}, \, \Phi^{\phantom *}_{b \vect{\omega}_1}, \, \Phi^{(\sigma_1)*}_{\vect\alpha_1 +b \vect{h}_i ,\vect{\bar\alpha}_1 + b \vect{h}_i})}
  {C(\Phi^{(\sigma_1)}_{\vect\alpha_1,\vect{\bar\alpha}_1}, \, \Phi^{\phantom *}_{b \vect{\omega}_1}, \, \Phi^{(\sigma_1)*}_{\vect\alpha_1 +b \vect{h}_j ,\vect{\bar\alpha}_1 + b \vect{h}_j})}
  \times
  \frac{C(\Phi^{(\sigma_1)}_{\vect\alpha_1 +b \vect{h}_i ,\vect{\bar\alpha}_1 + b \vect{h}_i}, \Phi^{(\sigma_2)}_{\vect\alpha_2,\vect{\bar\alpha}_2}, \wt\Phi_{\vect\alpha_3,\vect{\bar\alpha}_3})}
  {C(\Phi^{(\sigma_1)}_{\vect\alpha_1 +b \vect{h}_j,\vect{\bar\alpha}_1 + b \vect{h}_j}, \Phi^{(\sigma_2)}_{\vect\alpha_2,\vect{\bar\alpha}_2}, \wt\Phi_{\vect\alpha_3,\vect{\bar\alpha}_3})} \nn \\
  & \quad =   \left[\prod_{k \neq i} \frac{ \Gamma(A_k - A_i)}{\Gamma(1 - \overline{A}_k + \overline{A}_i)}
  \prod_{k \neq j}\frac{\Gamma(1 - \overline{A}_k + \overline{A}_j)}{ \Gamma(A_k - A_j)} \right] \nn \\
  & \qquad \qquad \times \left[ \frac{\Gamma(A_i+B_\ell) \Gamma(1 - \overline{A}_j - \overline{B}_\ell)}
       {\Gamma(A_j+B_\ell) \Gamma(1 - \overline{A}_i- \overline{B}_\ell)}
       \prod_{k \neq \ell} \frac{\Gamma(\overline{A}_i + \overline{B}_k)\Gamma(1 - A_j- B_k)}
            {\Gamma(\overline{A}_j+\overline{B}_k)\Gamma(1 - A_i- B_k)} \right] \,. \label{eq:shift}
\end{align}
The first braket in the right-hand side only depends on $\vect\alpha_1$ and $\vect{\bar\alpha}_1$.
Let us rewrite the second bracket as:
\begin{align*}
  \frac{\sin\pi(\overline{A}_i+\overline{B}_\ell) \sin\pi(A_j + B_\ell)}
       {\sin\pi(A_i + B_\ell) \sin\pi(\overline{A}_j + \overline{B}_\ell)}
       \prod_{k=1}^n \frac{\Gamma( \overline{A}_i+\overline{B}_k)\Gamma(1 - A_j- B_k)}
            {\Gamma(\overline{A}_j+\overline{B}_k)\Gamma(1 - A_i- B_k)} \,.
\end{align*}
We note that \eqref{eq:MXM=Y} is invariant under the exchange $(A_i,B_i) \leftrightarrow (\overline{A}_i,\overline{B}_i)$. Hence, up to a sign, one can replace the right-hand side of~\eqref{eq:shift} by the geometric mean:
\begin{equation} \label{eq:shift2}
  \frac{C(\Phi^{(\sigma_1)}_{\vect\alpha_1 +b \vect{h}_i ,\vect{\bar\alpha}_1 + b \vect{h}_i}, \Phi^{(\sigma_2)}_{\vect\alpha_2,\vect{\bar\alpha}_2}, \wt\Phi_{\vect\alpha_3,\vect{\bar\alpha}_3})}
       {C(\Phi^{(\sigma_1)}_{\vect\alpha_1 +b \vect{h}_j,\vect{\bar\alpha}_1 + b \vect{h}_j}, \Phi^{(\sigma_2)}_{\vect\alpha_2,\vect{\bar\alpha}_2}, \wt\Phi_{\vect\alpha_3,\vect{\bar\alpha}_3})}
       = \frac{K_i^{(\sigma_1)}(\vect\alpha_1,\vect{\bar\alpha}_1)}{K_j^{(\sigma_1)}(\vect\alpha_1,\vect{\bar\alpha}_1)}
       \times \sqrt{\prod_{k=1}^n \frac{\gamma(A_i+B_k)\gamma(\overline{A}_i+\overline{B}_k)}
            {\gamma(A_j+B_k)\gamma(\overline{A}_j+\overline{B}_k)}} \,,
\end{equation}
where $\gamma(x)=\Gamma(x)/\Gamma(1-x)$, and $K_i^{(\sigma_1)}(\vect\alpha_1,\vect{\bar\alpha}_1)$ and $K_j^{(\sigma_1)}(\vect\alpha_1,\vect{\bar\alpha}_1)$ are some normalisation factors.

\subsubsection{Shift equation from the null descendant of $\Phi_{-\vect\omega_1/b}$}

Let us replace the degenerate field $b \vect{\omega}_1$ by $-\vect{\omega}_1/b$ in the four-point function \eqref{eq:non-diagonal_correlation_function}:
\begin{equation*}
  \wh{\mathcal{G}}(z,\bar{z}) = \aver{\Phi^{(\sigma_2)}_{\vect{\alpha}_2^* , \vect{\bar\alpha}_2^*}(\infty) \Phi_{-\vect{\omega}_1/b}(z,\bar{z}) \wt\Phi_{\vect \alpha_3, \vect{\bar\alpha}_3}(1) \Phi^{(\sigma_1)}_{\vect{\alpha}_1 , \vect{\bar\alpha}_1}(0)} \,,
\end{equation*}
and demand that the monodromy exponents satisfy:
\begin{equation*}
  \wh\eta(\vect\alpha_1, \vect{\bar\alpha}_1) + \wh\eta(\vect\alpha_2, \vect{\bar\alpha}_2) + \wh\eta(\vect\alpha_3, \vect{\bar\alpha}_3) \in \mathbb{Z} \,.
\end{equation*}
One can write the decomposition:
\begin{equation*}
  \wh{\mathcal{G}}(z,\bar{z}) = \sum_{i,j} \wh X_{ij} \, \wh{F}_i(z) \wh{\overline F}_j(\bar z)
  = \sum_{k,\ell} \wh Y_{k\ell} \, \wh{G}_k(z) \wh{\overline G}_\ell(\bar z) \,,
\end{equation*}
where $(\wh{F}_i, \wh{\overline F}_j, \wh{G}_k,\wh{\overline G}_\ell)$ are the analogs of $(F_i, \overline F_j, G_k, \overline G_\ell)$ with $(A_i,\overline A_j, B_k, \overline B_\ell)$ replaced by:
\begin{align*}
  C_i &= \Delta_{\vect{\alpha}_1 - \vect{h}_i/b}  - \Delta_{\vect{\alpha}_1} - \mu/b \,, \quad
  &&\overline{C}_j = \Delta_{\vect{\bar\alpha}_1 - \vect{h}_j/b}  - \Delta_{\vect{\bar\alpha}_1} - \overline\mu/b \,, \\
  D_k &= \Delta_{\vect{\alpha}_2 - \vect{h}_k/b} - \Delta_{\vect{\alpha}_2} - \mu/b \,, \quad
  &&\overline{D}_\ell = \Delta_{\vect{\bar\alpha}_2 - \vect{h}_\ell/b}  - \Delta_{\vect{\bar\alpha}_2} - \overline\mu/b \,.
\end{align*}
Using the fusion rules~\eqref{eq:fusion2}, the coefficient matrices must be of the form:
\begin{equation*}
  \wh X_{ij} = \delta_{i,\sigma_1(j)} \, \wh X_j \,,
  \qquad \wh Y_{k\ell} = \delta_{k,\sigma_2(\ell)} \, \wh Y_\ell \,.
\end{equation*}
Moreover, these matrices are related by
\begin{equation*}
  \wh Y = N^t \, \wh X \, \overline{N} \,,
\end{equation*}
where $N=M(\{C_i\},\{D_j\})$ and $\overline{N}=M(\{\overline{C}_i\},\{\overline{D}_j\})$ in \eqref{eq:change_basis}. We get a relation similar to~\eqref{eq:MXM=Y}:
\begin{equation*}
  \sum_j N'_{jk} \, \wh X_j \, \overline{N}_{j\ell} = \delta_{k\ell} \, \wh Y_\ell \,,
\end{equation*}
where we have defined the matrix elements $N'_{jk} = N_{\sigma_1(j),\sigma_2(k)}$. This corresponds to the matrix in \eqref{eq:change_basis}:
\begin{equation*}
  N'=M(\{C'_i\},\{D'_j\}) \,, \qquad C'_i=C_{\sigma_1(i)} \,, \quad D'_j=D_{\sigma_2(j)} \,.
\end{equation*}
Reasoning as above, we obtain a shift equation analogous to~\eqref{eq:shift2}:
\begin{equation} \label{eq:shift3}
  \frac{C(\Phi^{(\sigma_1)}_{\vect\alpha_1 - \vect{h}_{\sigma_1(i)}/b,\vect{\bar\alpha}_1 -\vect{h}_i/b}, \Phi^{(\sigma_2)}_{\vect\alpha_2,\vect{\bar\alpha}_2}, \wt\Phi_{\vect\alpha_3,\vect{\bar\alpha}_3})}
       {C(\Phi^{(\sigma_1)}_{\vect\alpha_1 - \vect{h}_{\sigma_1(j)}/b, \vect{\bar\alpha}_1 -\vect{h}_j/b}, \Phi^{(\sigma_2)}_{\vect\alpha_2,\vect{\bar\alpha}_2}, \wt\Phi_{\vect\alpha_3,\vect{\bar\alpha}_3})}
       = \frac{\wh K_i^{(\sigma_1)}(\vect\alpha_1,\vect{\bar\alpha}_1)}{\wh K_j^{(\sigma_1)}(\vect\alpha_1,\vect{\bar\alpha}_1)}
       \times \sqrt{\prod_{k=1}^n \frac{\gamma(C_{\sigma_1(i)}+D_k)\gamma(\overline{C}_i+\overline{D}_k)}
            {\gamma(C_{\sigma_1(j)}+D_k)\gamma(\overline{C}_j+\overline{D}_k)}} \,,
\end{equation}
where $\wh K_i^{(\sigma_1)}(\vect\alpha_1,\vect{\bar\alpha}_1)$ and $\wh K_j^{(\sigma_1)}(\vect\alpha_1,\vect{\bar\alpha}_1)$ are some normalising factors.

\subsubsection{Solution of the shift equations}

The shift equations (\ref{eq:shift2}--\ref{eq:shift3}) have a form very close to the one for scalar operators \eqref{eq:shift-scalar}. Up to normalising factors, the right-hand side of these equations is simply the geometric mean of the right-hand side of \eqref{eq:shift-scalar}, with charges $(\vect\alpha_1,\vect\alpha_2,\vect\alpha_3)$ and $(\vect{\bar\alpha_1},\vect{\bar\alpha}_2,\vect{\bar\alpha}_3)$, respectively. The major difference with scalar operators is the fact that the constraints~\eqref{eq:constraints} impose a quantisation of the vertex charges $\vect\alpha$ and $\vect{\bar\alpha}$.

However, if one of the operators is scalar (say if $\sigma_1=\id$) then its vertex charge can take continuous values, and the solution takes the form:
\begin{equation} \label{eq:sol}
  C(\Phi_{\vect\alpha_1}, \Phi^{(\sigma_2)}_{\vect\alpha_2,\vect{\bar\alpha}_2}, \wt\Phi_{\vect\alpha_3,\vect{\bar\alpha}_3}) = \sqrt{C(\vect\alpha_1,\vect\alpha_2,\vect\alpha_3) \, C(\vect\alpha_1,\vect{\bar\alpha}_2,\vect{\bar\alpha}_3)} \,,
\end{equation}
where $C(\vect\alpha_1,\vect\alpha_1,\vect\alpha_3)$ is the structure constant of scalar operators, given in (\ref{eq:C-vertex}--\ref{eq:M}). Note that this result is valid only when $\Phi^{(\sigma_1)}_{\vect\alpha_1,\vect{\bar\alpha}_1}$ and $\Phi^{(\sigma_2)}_{\vect\alpha_2,\vect{\bar\alpha}_2}$ are non-degenerate, $\wt\Phi_{\vect\alpha_3,\vect{\bar\alpha}_3}$ is semidegenerate, and the $\mathbb{Z}_n$ charge neutrality conditions are satisfied:
\begin{equation}
  e^{2i\pi[\eta(\vect\alpha_1,\vect{\bar\alpha}_1) +\eta(\vect\alpha_2,\vect{\bar\alpha}_2) + \eta(\vect\alpha_3,\vect{\bar\alpha}_3)]} = 1 \,,
  \qquad e^{2i\pi[\wh\eta(\vect\alpha_1,\vect{\bar\alpha}_1) +\wh\eta(\vect\alpha_2,\vect{\bar\alpha}_2) + \wh\eta(\vect\alpha_3,\vect{\bar\alpha}_3)]} = 1 \,,
\end{equation}
where $\eta$ and $\wh\eta$ are defined in~\eqref{eq:eta}.

In the case of generic non-scalar operators, the vertex charges obey the quantisation conditions~\eqref{eq:constraints}, and the structure constants $C(\Phi^{(\sigma_1)}_{\vect\alpha_1,\vect{\bar\alpha}_1}, \Phi^{(\sigma_2)}_{\vect\alpha_2,\vect{\bar\alpha}_2}, \wt\Phi_{\vect\alpha_3,\vect{\bar\alpha}_3})$ are determined by the shift equations (\ref{eq:shift2}--\ref{eq:shift3}), up to an overall factor.

\section{Conclusion}

In this paper generic $W_n$ symmetric CFTs are considered. Using the analytic conformal bootstrap, a class of three point functions in the imaginary $\mathfrak{sl}_n$ Toda field theory is computed. As in the case of a real background charge, these results are restricted to three-point functions involving two arbitrary (scalar) fields and one semi-degenerate field of Wyllard type.  Non-scalar primary fields are also considered. Imposing a well-defined monodromy with the fully-degenerate fields $\Phi_{b \vect{\omega}_1}$ and $\Phi_{-b^{-1} \vect{\omega}_1}$ leads to a classification of non-scalar fields by conjugacy classes of the permutation group $\mathfrak{S}_n$.  The conformal bootstrap is extended to include these non-scalar fields, and the corresponding shift equations obeyed by the structure constantes are obtained. Three-point functions involving two arbitrary fields and one semi-degenerate field of Wyllard type (possibly non-scalar) are computed explicitly as long as one of the generic fields is scalar.

\section*{Acknowledgements}

The authors wish to thank Sylvain Ribault and Santiago Migliaccio for insightful discussions.

\appendix 

\section{Conventions for \texorpdfstring{$\mathfrak{sl}_n$}{sl(n)} : roots, weights and Weyl group}
\label{app:sln_conventions}

Many features of the Toda field theory are related to the $\mathfrak{sl}_n$ Lie algebra. We recall basic facts and notations in this appendix.

\subsection{Conventions for \texorpdfstring{$\mathfrak{sl}_3$}{sl(3)} }

The Lie algebra $\mathfrak{sl}_3$ has two simple roots $\vect{e_1}$ and $\vect{e_2}$, its Cartan matrix, defined by the scalar product $K_{i,j} = \vect{e}_i \cdot \vect{e}_j$, takes the form:
\[
K = \begin{pmatrix}
2 &-1 \\
-1 & 2 \\
\end{pmatrix} \,.
\]
The weights $\vect{\omega}_i$ of the Lie algebra are dual to its roots, $\vect{e}_i \cdot \vect{\omega}_j = \delta_{i,j}$. They can be written:
\[
\vect{\omega}_1 = \frac{1}{3} \left(2 \vect{e}_1 + \vect{e}_2\right), \ \  \vect{\omega}_2 = \frac{1}{3} \left(2 \vect{e}_2 + \vect{e}_1\right) \quad \Rightarrow \quad \vect{\omega_1} \cdot \vect{\omega_1} = \vect{\omega_2} \cdot \vect{\omega_2}  = \frac{2}{3} \text{ and } \vect{\omega_1} \cdot\vect{\omega_2} = \frac{1}{3}
\]
The weights of the first fundamental representation are defined as:
\[ 
\vect{h}_{1} = \vect{\omega_1}, \qquad  \vect{h}_{2} = \vect{\omega}_1 - \vect{e}_1 = \vect{\omega}_2 - \vect{\omega}_1, \qquad \vect{h}_{3} = \vect{\omega}_1 - \vect{e}_1 - \vect{e}_2 = -\vect{\omega}_2
\] 
The root lattice is $\mathcal{R} = \mathbb{Z}\vect{e}_1 + \mathbb{Z}\vect{e}_2$, and its dual $\mathcal{R}^* = \mathbb{Z}\vect\omega_1 + \mathbb{Z}\vect\omega_2$ is the weight lattice.
The Weyl vector can be written both in terms of $\vect{e}_i$ and $\vect{\omega}_j$: $\vect{\rho} = \vect{\omega}_1 + \vect{\omega}_2 = \vect{e}_1 + \vect{e}_2 = \vect{h}_1 - \vect{h}_3$. 

\begin{figure}
  \centering
  \includegraphics[width=.5\linewidth]{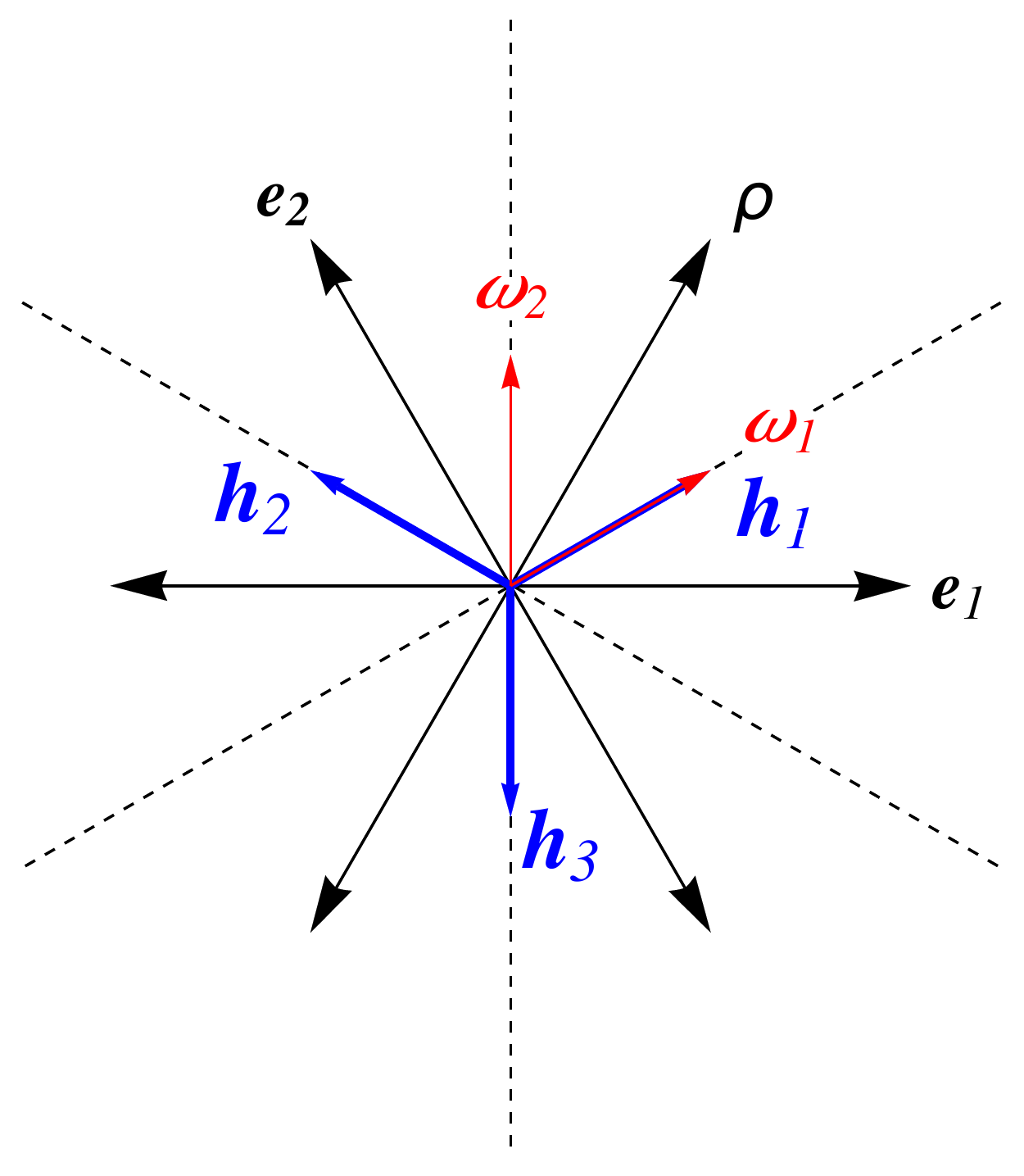}
  \caption{The generators of the root lattice $\mathcal{R}$ (in black) and the weight lattice $\mathcal{R}^*$ (in blue/red) for $\mathfrak{sl}_3$. The reflections of the Weyl group are the reflections \emph{w.r.t.} the dashed lines, while the rotation $R_j$ ($j=1,2,3$) sends $\vect{h}_1$ to $\vect{h}_j$. }
    \label{Fig1}
\end{figure}

The Weyl group $\cal W$ is generated by the reflections $s_1$ and $s_2$ 
\[
s_i ( \vect{v}) = \vect{v} - (\vect{v} \cdot \vect{e}_i) \vect{e}_i \,.
\]
It has $6$ elements, three reflections $s_j$ and three rotations $R_j$ (see Fig. \ref{Fig1}), $j=1,2,3$. In terms of the generators $s_1$ and $s_2$ one has 
\begin{align*}
R_1 = 1, \qquad R_2 = s_1s_2, \qquad R_3 = s_2 s_1, \qquad s_3 = s_1s_2s_1 = s_2 s_1 s_2
\end{align*}
The Weyl group of $\mathfrak{sl}_3$ can be identified with $\mathfrak{S}_3$, the group of permutations of three elements. In terms of the $\vect{h}_{i}$, the Weyl group acts by permutations $\vect{h}_{i}\ \to \vect{h}_{\sigma(i)}$. In the following we will denote an element of the Weyl group by the corresponding permutation $\sigma$.

Finally, the duality operation $\vect{\alpha} \mapsto \vect{\alpha^{*}}$ is the reflection with respect to $\vect{\rho}$ (this is the unique linear involution exchanging $\vect{\omega}_1$ and $\vect{\omega}_2$). This reflection does not belong to the Weyl group. 

\subsection{Conventions for \texorpdfstring{$\mathfrak{sl}_n$}{sl(n)}} 

The generalization to $\mathfrak{sl}_n$ is straightforward. Let $\{ \vect{h}_i, \,i = 1,\cdots,n \}$ be the first fundamental representation of $\mathfrak{sl}_n$, normalized as
\begin{align}
  \vect{h}_i \cdot \vect{h}_j = \delta_{ij} - \frac{1}{n} \,. 
\end{align}
The root and weight lattices are 
\begin{align}
  \mathcal{R} = \sum_{i=1}^n \mathbb{Z} \vect{e}_i \,,
  \qquad \mathcal{R}^* = \sum_{i=1}^n \mathbb{Z} \vect{\omega}_i \,,
\end{align}
where the fundamental weights $\vect{\omega}_i$ and the simple roots $\vect{e}_i$ can be expressed as 
\begin{align*}
  \vect{e}_i = \vect{h}_i - \vect{h}_{i+1}  \,,
  \qquad \vect\omega_i  = \sum_{k=1}^i \vect{h}_k \,,
\end{align*}
for $i = 1, \cdots, n-1$. The Weyl vector is 
\begin{align}
  \vect{\rho} = \sum_{i=1}^{n-1} \vect{\omega}_i   = - \sum_{i=1}^{n} i \vect{h}_i  \,,
  \qquad \vect{\rho} \cdot \vect{\rho} = \frac{n(n^2 - 1)}{12} \,.
\end{align}
The Weyl group is isomorphic to $\mathfrak{S}_n$, and acts as $\sigma (\vect{h}_i) = \vect{h}_{\sigma(i)}$. In particular the reflection $s_i : \vect{x} \to  \vect{x} - (\vect{e}_i \cdot \vect{x}) \vect{e}_i$ is mapped to the transposition $\tau_{i,i+1}$, and the longest element of Weyl group (denoted by $s_0$), which for $\mathfrak{sl}_n$ reads 
\begin{align}
  s_0 = s_1 s_2 \cdots s_{n-1}   s_1 s_2 \cdots s_{n-2}  \cdots s_1 s_2 s_1
\end{align}
corresponds to the permutation
\begin{align}
  \rho (i) = n-i\, .
\end{align}
Finally, the dual operation is defined as
\begin{align}
  \vect{x}^* = - s_0 (\vect{x})\, .
\end{align}

\section{Hypergeometric conformal blocks}
\label{app:hypergeometric}

\subsection{Hypergeometric conformal blocks for \texorpdfstring{$\mathfrak{sl}_3$}{sl(3)}}
\label{app:hypergeometric_sl3}

Event though the fusion of the form $\Phi_{b \vect{\omega}_1} \otimes \Phi_{\vect\alpha}$ only gives rise to three primary operators as in \eqref{eq:fusion_rule_degenerate}, it is known that a generic four point function
$\aver{\Phi_{b\vect{\vect{\omega}_1}}\Phi_{\vect{\alpha}_1} \Phi_{\vect{\alpha}_2} \Phi_{\vect{\alpha}_3}}$
has more than three Virasoro conformal blocks. However, if one of the fields $\vect\alpha_i$ is semi-degenerate, a third-order differential equation can be obtained for this function \cite{FL1,FL2,Belavin:2016qaa,Belavin2017,Belavin2017b}. The simplest semi-degenerate fields correspond to $\vect\alpha_3=\kappa \vect\omega_j$ ($j=1,2$), with a null-vector at level $1$ as in \eqref{eq:W3_semi_degenerate_nve}. It was found in Fateev and Litvinov for the real Toda theory \cite{FL1,FL2} that the correlation function 
\[
\mathcal{G}(z) =  \langle \Phi_{\vect{\beta}^{*}} | \Phi_{b \vect{\omega}_1}(z)  \Phi_{\kappa \vect{\omega}_j}(1)  | \Phi_{\vect{\alpha}}\rangle
\]
obeys a Fuchsian differential equation of order $3$, whose solutions are given in terms of generalised hypergeometric functions. In this section we adapt these resuts in the case of imaginary Toda.  

The residue theorem applied to the function 
\[
w \rightarrow \langle \Phi_{\vect{\beta}^{*}} | W(w) \Phi_{b \vect{\omega}_1}(z)  \Phi_{\kappa \vect{\omega}_j}(1)  | \Phi_{\vect{\alpha}}\rangle \frac{(w-1) w^2}{w-z} 
\]
yields
\begin{equation*}
  \begin{split}
0 = &\left( w_{\vect{\beta}} + w_{b \vect{\omega}_1} + \frac{w_{\vect{\alpha}}}{z} + \frac{ (1 - 2z)}{(z-1)^2} w_{\kappa \vect{\omega}_j}\right)  \langle \Phi_{\vect{\beta}^{*}} | \Phi_{b \vect{\omega}_1}(z)  \Phi_{\kappa \vect{\omega}_j}(1)  | \Phi_{\vect{\alpha}}\rangle \\
 +& \  \frac{1}{1-z} \langle \Phi_{\vect{\beta}^{*}} | \Phi_{b \vect{\omega}_1}(z) \left( W_{-1} \Phi_{\kappa \vect{\omega}_j} \right)(1)  | \Phi_{\vect{\alpha}}\rangle 
 + (3z-1) \langle \Phi_{\vect{\beta}^{*}} | \left(W_{-1}  \Phi_{b \vect{\omega}_1} \right)(z)  \Phi_{\kappa \vect{\omega}_j}(1)  | \Phi_{\vect{\alpha}}\rangle \\ + & \ z (3 z - 2)  \langle \Phi_{\vect{\beta}^{*}} | \left( W_{-2} \Phi_{b \vect{\omega}_1} \right)(z)  \Phi_{\kappa \vect{\omega}_j}(1)  | \Phi_{\vect{\alpha}}\rangle + z^2(z-1) \langle \Phi_{\vect{\beta}^{*}} | \left( W_{-3} \Phi_{b \vect{\omega}_1} \right)(z)  \Phi_{\kappa \vect{\omega}_j}(1)  | \Phi_{\vect{\alpha}}\rangle \,.
  \end{split}
\end{equation*}
By using the null-vector equations \eqref{eq:null_vector_W3}, it is possible to rewrite this equation in terms Virasoro modes, leading to a differential equation for the correlation function. This differential equation takes a very simple form in term of the function
\[
f(z) := \mathcal{G}(z)  (1-z)^{- 2b\mu} z^{b\mu + \Delta_{b \vect{\omega}_1}},
\]
where 
\begin{equation*}
\mu = 
\begin{cases}
 \frac{1}{2}\kappa \vect{\omega}_2 \cdot \vect{h}_1 \quad  & \text{ if }  \  j=2 \\
   \frac{1}{2}\left( \vect{Q} - \kappa \vect{\omega}_2 \right) \cdot \vect{h}_1 \quad & \text{ if }  \ j=1 \\
\end{cases}
\end{equation*}
The function $f(z)$ obeys the following Fuchsian differential equation
\begin{equation} \label{eq:differential_equation}
z\left(D + B_1  \right)\left(D + B_2 \right)\left(D +
      B_3 \right)f(z) = \left(D -A_1\right)\left(D -A_2\right) \left(D - A_3\right) f(z).
\end{equation}
where $D = z \diff{}{z}$ and
\begin{equation}
 A_i = \Delta_{\vect{\alpha} + b \vect{h}_i}  - \Delta_{\vect{\alpha}}+  b\mu \,,
 \qquad B_i =  \Delta_{\vect{\beta} + b \vect{h}_i}  -\Delta_{\vect{\beta}}  + b\mu \,.
\end{equation}
Note that the equation is invariant under the simultaneous change $z \to z^{-1}$ and $A_i \leftrightarrow B_i$. This simply reflects the fact that
\begin{align*}
\langle \Phi_{\vect{\beta}^{*}} | \Phi_{\vect{b \vect{\omega}_1}}(z)  \Phi_{\kappa \vect{\omega}_j}(1)  | \Phi_{\vect{\alpha}}\rangle = \langle \Phi_{\vect{\alpha}^{*}} | \Phi_{\vect{b \vect{\omega}_1}}(1/z)  \Phi_{\kappa \vect{\omega}_j}(1)  | \Phi_{\vect{\beta}}\rangle  z^{-2 \Delta_{b \vect{\omega}_1}} \,.
\end{align*}
The Riemann scheme of this Fuchsian differential equation is
\begin{equation}
  \label{Riem_not_app}
  \begin{Bmatrix}  
    z=0& z=1 & z=\infty  \\ 
    \hline
    A_1  & 0  & B_1    \\
    A_2  &  1  & B_2   \\
    A_3  &  2-\sum_i (A_i + B_i)    & B_3  
  \end{Bmatrix}
\end{equation}
and the sum of all exponents is $3$, as it should according to the Fuchs relation. The exponents as $z \to1$ are compatible with the fusion rules \eqref{eq:fusion_rules_degenerate_semi_degenerate}. 
\bigskip

A basis of solutions with Abelian monodromies around $z=0$ is given by
\begin{align*}
  f_i(z) = (-z)^{A_i} \ \pFq{3}{2}{B_1 + A_i, \cdots,  B_3+ A_i }{1- A_1 + A_i, \cdot \cdot {}^* \cdot \cdot,  1- A_3+ A_i}{z}\,,
\end{align*}
where $\cdot \cdot {}^* \cdot \cdot$ denotes suppression of the term $1 - A_i + A_i$. Likewise, the solutions will Abelian monodromies around $\infty$ are simply obtained through $A_i \leftrightarrow B_i$ and $z \to z^{-1}$ :
\begin{align*}
  g_i(z) = (-z)^{-B_i} \ \pFq{3}{2}{A_1 + B_i, \cdots,  A_3+ B_i }{1- B_1 + B_i, \cdot \cdot {}^* \cdot \cdot,  1- B_3+ B_i}{\frac{1}{z}}\,.
\end{align*}
Going back to the function $\mathcal{G}$, we have the following conformal blocks
\begin{align}
F_i(z) & = (1-z)^{2b\mu}\ (-z)^{\eta_i} \ \pFq{3}{2}{B_1 + A_i, \cdots,  B_3+ A_i }{1- A_1 + A_i, \cdot \cdot {}^* \cdot \cdot,  1- A_3+ A_i}{z}\,, \\
 G_i(z) & =  \left(1-\frac{1}{z}\right)^{2 b\mu}  \left(-\frac{1}{z} \right)^{ \zeta_i}  \pFq{3}{2}{A_1 + B_i, \cdots,  A_3+ B_i }{1- B_1 + B_i, \cdot \cdot {}^* \cdot \cdot,  1- B_3+ B_i}{\frac{1}{z}}\,,
\label{eq:solutions}
\end{align}
where
\begin{align*}
\eta_i & = A_i - b\mu - \Delta_{b \vect{\omega}_1} = \Delta_{\vect{\alpha} + b \vect{h}_i}  - \Delta_{\vect{\alpha}}- \Delta_{b \vect{\omega}_1} \\
\zeta_i & = B_i - b\mu + \Delta_{b \vect{\omega}_1}  =  \Delta_{\vect{\beta} + b \vect{h}_i}  -\Delta_{\vect{\beta}} + \Delta_{b \vect{\omega}_1} 
\end{align*}
are the fusion exponents as $z \to 0$ and $z\to \infty$. 
\bigskip

Acting with the Weyl group on $\vect{\alpha}$ simply permutes the blocks as follows
\begin{align*}
  F_i \to F_{\sigma(i)}, \qquad G_i \to G_i
\end{align*}
while reparametrization of $\vect{\beta}$ yields
\begin{align*}
  F_i \to F_{i}, \qquad G_i \to G_{\sigma(i)}
\end{align*}

The two bases are related through $F_i(z) = M_{ij} G_j(z)$ :
\begin{align*}
  M_{ij} = \prod_{k \neq i} \frac{\Gamma(1 + A_i -A_k)}{\Gamma(1 - B_j -A_k)}
  \, \prod_{\ell \neq j} \frac{\Gamma(B_l-B_j)}{\Gamma(B_\ell +A_i)}
  = \prod_{k \neq i}\frac{\Gamma(1+ \eta_i  - \eta_k)}{\Gamma(1 - 2b\mu - \zeta_j - \eta_k)}
  \, \prod_{\ell \neq j} \frac{\Gamma(\zeta_\ell - \zeta_j)}{\Gamma(2b\mu + \zeta_\ell + \eta_i)} \,.
\end{align*}
The coefficients of $M^{-1}$ are obtained by exchanging $A_i \leftrightarrow B_i$. 

\subsection{Hypergeometric conformal blocks for \texorpdfstring{$\mathfrak{sl}_n$}{sl(n)}}

The generalization to $\mathfrak{sl}_n$ is as follows. Consider the correlation function
\[
f(z) :=\langle \Phi_{\vect{\beta}^{*}} | \Phi_{b \vect{\omega}_1}(z)  \Phi_{\kappa \vect{\omega}_j}(1)  | \Phi_{\vect{\alpha}}\rangle\, (1-z)^{- 2b\mu} z^{b\mu + \Delta_{b \vect{\omega}_1}},
\]
where $j =1$ or $j=n-1$, and 
\begin{equation*}
  2 b\mu = 
  \begin{cases}
    \Delta_{\kappa \vect{\omega}_2 + b \vect{h}_1} -  \Delta_{\kappa \vect{\omega}_2}  - \Delta_{b \vect{\omega}_1}   =   \frac{b \kappa}{n}  \quad  & \text{ if }  \  j=n-1 \\
    \Delta_{\kappa \vect{\omega}_1 + b \vect{h}_2} -  \Delta_{\kappa \vect{\omega}_1}  - \Delta_{b \vect{\omega}_1}   =  -\frac{b \kappa}{n} + (1 - b^2) \quad & \text{ if }  \ j=1 \\
  \end{cases}
\end{equation*}
The function $f(z)$ obeys the following Fuchsian differential equation
\begin{equation} 
  z\left(D + B_1  \right)\left(D + B_2 \right)\cdots \left(D +
  B_n \right)f(z) = \left(D -A_1\right)\left(D -A_2\right) \cdots  \left(D - A_n\right) f(z)
\end{equation}
where 
\begin{equation}
  A_i = \Delta_{\vect{\alpha} + b \vect{h}_i}  - \Delta_{\vect{\alpha}} + b\mu \,,
  \qquad B_i =  \Delta_{\vect{\beta} + b \vect{h}_i}  -\Delta_{\vect{\beta}} + b\mu \,.
\end{equation}
The Riemann scheme is
\begin{equation} \label{Riem_not}
  \begin{Bmatrix}  
    z=0    & z=1                  & z=\infty  \\
    \hline
    A_1    & 0                     & B_1    \\
    A_2    & 1                     & B_2   \\
    \vdots & \vdots                & \vdots   \\
    A_{n-1} & n-2                   & B_{n-1}   \\
    A_n    & n-1-\sum_i (A_i + B_i) & B_n  
  \end{Bmatrix}
\end{equation}
and the exponents as $z\to 1$ give the fusion rules
\begin{equation}
  \label{eq:fusion_rules_degenerate_semi_degenerate_sln}
  \Phi_{b \omega_1} \otimes  \Phi_{\kappa \omega_{n-1}} \rightarrow
  \Phi_{ \kappa \omega_{n-1} + b \vect{h}_1 }
  \oplus \Phi_{ \kappa \omega_{n-1} + b \vect{h}_n} \,,
  \quad  \Phi_{b \omega_1} \otimes  \Phi_{\kappa \omega_1} \rightarrow
  \Phi_{ \kappa \omega_1 + b \vect{h}_1 } \oplus \Phi_{ \kappa \omega_1 + b \vect{h}_2}\,.
\end{equation}
A basis of solutions can be obtained by series expansion around $z=0$, namely
\begin{align*}
f_i(z) = (-z)^{A_i} \ \pFq{n}{n-1}{B_1 + A_i, \cdots,  B_n+ A_i }{1- A_1 + A_i, \cdot \cdot {}^* \cdot \cdot,  1- A_n+ A_i}{z}\,,
\end{align*}
where $\cdot \cdot {}^* \cdot \cdot$ denotes suppression of the term $1 - A_i + A_i$.  The above series is convergent for $|z| <1$, and it can be analytically continued. The Weyl group (reparametrization of $\vect\alpha$) acts by permutations on these $n$ conformal blocks. 

Likewise, around $z=\infty$ the solutions with Abelian monodromies around $\infty$ are simply obtained through $A_i \leftrightarrow B_i$ and $z \to z^{-1}$ :
\begin{align*}
  g_i(z) = (-z)^{-B_i} \ \pFq{n}{n-1}{A_1 + B_i, \cdots,  A_n+ B_i }{1- B_1 + B_i, \cdot \cdot {}^* \cdot \cdot,  1- B_n+ B_i}{\frac{1}{z}}\,.
\end{align*} 
The change of bases
\begin{align*}
  f_i(z) = \sum_j \, M_{ij} \, g_j(z)
\end{align*}
can be obtained using contour deformation of the following Mellin-Barnes integral
\begin{align*}
  I(z) = \frac{1}{2\pi i} \int ds\, \Gamma(B_1 + s) \cdots \Gamma(B_n +s)\Gamma(A_1 - s) \cdots \Gamma(A_n - s)  \left( \epsilon z \right)^s \,,
\end{align*}
where $\epsilon = (-1)^n$ and the integration contour goes from $i \infty$ to $-i \infty$ while keeping all the poles $ \{A_i + k, \, k \in \mathbb{N} \}$ to the left and the poles  $ \{-B_i - k, \, k \in \mathbb{N} \}$ to the right. One finds
\begin{align}
  \label{eq:change_basis}
  M_{ij} = \prod_{k \neq i}  \frac{\Gamma(1 + A_i -A_k)}{\Gamma(1 - B_j -A_k)}
  \, \prod_{\ell \neq j}  \frac{\Gamma(B_\ell-B_j)}{\Gamma(B_\ell +A_i)}  \,,
\end{align}
and the coefficients of $M^{-1}$ are obtained by exchanging $A_i \leftrightarrow B_i$.

\section{Upsilon and double Gamma functions}
\label{app:Upsilon}

For $0 < \mathrm{Re}(x) < b + b^{-1}$, the function $x\mapsto\Upsilon_{b}(x)$ is given by:
\begin{equation} \label{ups}
  \ln \Upsilon_b(x)\equiv\!\!\int_0^\infty {\frac{\dif t}{t}}\!\!\left[\left({\frac{b + b^{-1}}{2}}-x\right)^2\!\! {\rm e}^{-t}-\frac{\sinh^2\left[\left(\frac{b + b^{-1}}{2}-x\right){\frac{t}{2}}\right]}{\sinh{\frac{b t}{2}}\sinh{\frac{t}{2 b}}}\right] \,.
\end{equation}
Outside of this interval, the function can be computed using the recursion formulas:
\begin{equation} \label{analyticcont}
  \begin{aligned}
    \Upsilon_b (x+b) &= \gamma(b x) \ b^{1-2 b x} \ \Upsilon_b(x) \,, \\
    \Upsilon_b(x+b^{-1}) &= 
    \gamma(x b^{-1}) \ b^{-1+2xb^{-1}} \ \Upsilon_b(x) \,. 
  \end{aligned}
\end{equation}
Moreover, it is clear from the integral definition that
\begin{equation} \label{Upsilon_properties}
  \begin{aligned}
   \Upsilon_b (x) &=  \Upsilon_{b^{-1}}(x) \,, \\
        \Upsilon_b(x) &= 
    \Upsilon_b (b+b^{-1} -x ) \,. 
  \end{aligned}
\end{equation}

$\Gamma_b = \Gamma_{b^{-1}}$ is a double Gamma function with periods $b$ and $b^{-1}$. It enjoys
\begin{equation} \label{analyticcont_gammab}
  \begin{aligned}
 \Gamma_b (x+b) &=  \sqrt{2\pi}\frac{b^{b x-1/2}}{\Gamma( b x )} \ \Gamma_b(x) \,, \\
    \Gamma_b(x+b^{-1}) &=  \sqrt{2\pi}
    \frac{b^{- x/b-1/2}}{\Gamma( x/b )} \ \Gamma_b(x)  \,. 
  \end{aligned}
\end{equation}
These two functions are related through 
\begin{align}
  \Upsilon_b(x) = \frac{1}{\Gamma_b (x) \Gamma_b ( b+b^{-1} -x)}
\end{align}

\section{Explicit charge lattices for non-scalar operators}

\subsection{The \texorpdfstring{$\mathfrak{sl}_2$}{sl(2)} case}

We take the conventions $e_1=\sqrt{2}$, and $h_1 = -h_2 = \omega_1 = 1/\sqrt 2$. The root lattice is then $\mathcal{R} = \mathbb{Z} \sqrt{2}$, and the weight lattice is $\mathcal{R}^*=\mathbb{Z}/\sqrt 2$. The background charge is given by $Q=(b^{-1}-b)/\sqrt{2}$. The central charge and conformal dimensions read
\begin{equation*}
  c = 1 - 6(b-b^{-1})^2 \,,
  \qquad \Delta_\alpha = \frac{1}{2} \alpha(\alpha-2Q) \,.
\end{equation*}
Scalar operators correspond to $\sigma= \id$, and have unconstrained vertex charges $\alpha=\bar\alpha$. Non-scalar operators $\Phi_{\alpha,\bar\alpha}^{(\sigma)}$ correspond to the transposition $\sigma=(12)$, which translates into the constraints:
\begin{equation*}
  \alpha-\bar\alpha \in \mathbb{Z}/(b\sqrt 2) \,,
  \qquad \alpha+\bar\alpha-2 Q \in b\mathbb{Z}/\sqrt 2 \,.
\end{equation*}
The solution is the set of charges:
\begin{equation*}
  \alpha = \frac{(1-r)b^{-1} - (1-s)b}{\sqrt 2} \,,
  \qquad \bar \alpha = \frac{(1+r)b^{-1} - (1-s)b}{\sqrt 2} \,,
  \qquad (r,s) \in (\mathbb{Z}/2)^2 \,.
\end{equation*}
These correspond to the conformal dimensions $(\Delta,\bar\Delta)=(\Delta_{rs},\Delta_{-r,s})$ of the Kac table, with \emph{half-integer} indices $r$ and $s$. The non-generic charges in the sense of \eqref{eq:alpha-generic} and \eqref{eq:alpha-generic2} correspond to the dimensions $\Delta_{k0}$ or $\Delta_{0k}$, with integer $k$.

\subsection{The \texorpdfstring{$\mathfrak{sl}_3$}{sl(3)} case}

The constraints \eqref{eq:constraints} may be written:
\begin{equation} \label{eq:constraints2}
  \begin{cases}
    (\id-\sigma)(\vect\alpha - \vect Q) &= b \vect{s} - b^{-1} \sigma \vect{r} \,, \\
    (\id-\sigma)(\vect{\bar\alpha} - \vect Q) &= b \vect{s} - b^{-1} \vect{r} \,,
  \end{cases}
  \qquad \text{with} \qquad (\vect{s},\vect{r}) \in (\mathcal{R^*})^2 \,.
\end{equation}
Let us discuss the allowed vertex charges in $\Phi_{\vect\alpha,\vect{\bar\alpha}}^{(\sigma)}$ for the various choices of conjugacy classes for $\sigma$:
\begin{itemize}
  
\item If $\sigma = \id$ we get $b \vect s = b^{-1} \vect r$, and thus $\vect r = \vect s=0$ since $b^2$ is irrational. This leaves $\vect\alpha$ unconstrained, and simply forces $\vect\alpha=\vect{\bar\alpha}$ : this corresponds to scalar operators.
  
\item If $\sigma$ is a cyclic permutation, {\it e.g.} $\sigma=(123)$, then $(\id-\sigma)$ is invertible, and we have $(\id-\sigma)^{-1} \mathcal{R}^* = \mathcal{R}/3$. We then get:
  \begin{equation*}
    \begin{cases}
      \vect\alpha &=  \vect Q + \frac{1}{3}(b \vect{m} - b^{-1} \sigma \vect{n}) \,, \\
      \vect{\bar\alpha} &= \vect Q + \frac{1}{3}(b \vect{m} - b^{-1} \vect{n}) \,,
    \end{cases}
    \qquad \text{with} \qquad (\vect{n},\vect{m}) \in \mathcal{R}^2 \,.
  \end{equation*}
  The vectors of $\mathcal{R}/3$ are of the form $\vect{m} = m_1 \vect\omega_1 + m_2 \vect\omega_2$, with $(m_1,m_2) \in \mathbb{Z}/3$, and $m_1-m_2 \in \mathbb{Z}$. Hence, we can write for $\sigma=(123)$:
  \begin{equation*}
    \vect\alpha = \vect\alpha \left(\begin{array}{cc} -n_1-n_2 & m_1 \\ n_2 & m_2 \end{array}\right) \,,
    \qquad \vect{\bar\alpha} = \vect\alpha \left(\begin{array}{cc} n_1 & m_1 \\ n_2 & m_2 \end{array}\right) \,,
  \end{equation*}
  where $(n_1,n_2,m_1,m_2) \in (\mathbb{Z}/3)^4$ satisfy $n_1-n_2 \in \mathbb{Z}$ and $m_1-m_2 \in \mathbb{Z}$, and we have used the notation for the charges in the Kac table \cite{Fateev87}:
  \begin{equation*}
    \vect\alpha \left(\begin{array}{cc} n_1 & m_1 \\ n_2 & m_2 \end{array}\right)
    = \left[(1-n_1)b^{-1} - (1-m_1)b \right] \vect\omega_1
    + \left[(1-n_2)b^{-1} - (1-m_2)b \right] \vect\omega_2 \,.
  \end{equation*}

\item If $\sigma$ is a transposition, {\it e.g.} $\sigma=(12)$, it is convenient to decompose the charge vectors on the orthogonal basis $(\vect{h}_3,\vect{e}_1)$, because $h_3$ is invariant under $\sigma$. Projecting the constraints \eqref{eq:constraints} onto $h_3$, one finds that
  $$(\vect\alpha-\vect{\bar\alpha}) \cdot \vect{h}_3 \in \frac{b^{-1}}{3} \mathbb{Z} \cap \frac{b}{3} \mathbb{Z} \,.$$
  Since $b^2$ is irrational, the above quantity vanishes, and hence we have:
  $$(\vect\alpha - \vect Q) \cdot h_3 = (\vect{\bar\alpha} - \vect Q) \cdot h_3 \,.$$
  The equations~\eqref{eq:constraints2} determine the components of $(\vect\alpha - \vect Q)$ and $(\vect{\bar\alpha} - \vect Q)$ on $e_1$. We find charges of the form:
  \begin{equation*}
    \begin{cases}
      \vect\alpha &= \vect{Q} + \beta \vect h_3 + \frac{1}{2}(-rb^{-1}+s b) \vect{e}_1 \,, \\
      \vect{\bar\alpha} &= \vect{Q} + \beta \vect h_3 + \frac{1}{2}(+rb^{-1}+s b) \vect{e}_1 \,,
    \end{cases}
    \qquad \text{with} \quad \beta \in \mathbb{R}, \quad (r,s) \in (\mathbb{Z}/2)^2 \,.
  \end{equation*}
  This result contains a mix of the features from the previous cases: the component on $\vect h_3$ is unconstrained, whereas the component on $\vect e_1$ has the same form as non-scalar charges in the $\mathfrak{sl}_2$ theory. If $r$ and $s$ are integers we are dealing with a semi-degenerate field with a null-vector at level $| r s|$. 
\end{itemize}

Let us finally comment on non-generic operators: if we write $\vect\alpha-\vect{Q} = x \omega_1 + y \omega_2$, then $\vect\alpha$ is non-generic as soon as one of the quantities $x$, $y$, or $(x+y)$ is an element of $b\mathbb{Z}$ or $\mathbb{Z}/b$. This includes, for instance, charges of the form
$$
\vect\alpha \left(\begin{array}{cc} 0 & k \\ \star & \star \end{array}\right) \,,
\qquad \vect\alpha \left(\begin{array}{cc} k & 0 \\ \star & \star \end{array}\right) \,,
\qquad \vect\alpha \left(\begin{array}{cc} \star & \star \\ 0 & k \end{array}\right) \,,
\qquad \vect\alpha \left(\begin{array}{cc} \star & \star \\ k & 0 \end{array}\right) \,,
$$
where $k$ is an integer, and the $\star$'s can take any independent real values.

\subsection{The \texorpdfstring{$\mathfrak{sl}_n$}{sl(n)} case}

Like for $\mathfrak{sl}_3$, the allowed vertex charges for an operator $\Phi_{\vect\alpha,\vect{\bar\alpha}}^{(\sigma)}$ may be described through the cycle decomposition of the permutation $\sigma$. Each cycle of length $m>1$ in $\sigma$ corresponds to an $m$-dimensional component of $(\vect\alpha,\vect{\bar\alpha})$ constrained to a lattice determined by an $\mathfrak{sl}_m$ theory. Each fixed point in $\sigma$ corresponds to an unconstrained one-dimensional component of $(\vect\alpha,\vect{\bar\alpha})$, with the same contribution to $\vect\alpha$ and $\vect{\bar\alpha}$.

The extremal cases are: (i) if $\sigma=\id$, then $\vect\alpha=\vect{\bar\alpha}$ with no further constraint; (ii) if $\sigma$ is a cyclic permutation of $n$ elements, then $(\vect\alpha,\vect{\bar\alpha})$ can be expressed in terms of two elements of the weight lattice $\mathcal R^*$ of $\mathfrak{sl}_n$.

\bibliographystyle{unsrt}
\bibliography{biblio}

\end{document}